\documentclass[twocolumn,prd,superscriptaddress,showpacs,floatfix,%
preprintnumbers,nofootinbib]{revtex4}

\usepackage{amssymb}

\usepackage{bm}
\usepackage{epsfig}
\usepackage{graphicx}

\usepackage{rotating}
\usepackage[dvips]{color}
\definecolor{Black}{named}{Black}
\definecolor{Red}{named}{Red}

\def\lsim{\raise0.3ex\hbox{$\;<$\kern-0.75em\raise-1.1ex\hbox{$\sim\;$}}}
\def\gsim{\raise0.3ex\hbox{$\;>$\kern-0.75em\raise-1.1ex\hbox{$\sim\;$}}}
\renewcommand{\baselinestretch}{1.3}

\def\theta{\vartheta}
\def\phi{\varphi}

\newcommand{\I}{{\rm i}}

\newcommand{\D}{{\rm d}}

\newcommand{\be}{\begin{equation}}
\newcommand{\ee}{\end{equation}}
\newcommand{\bea}{\begin{eqnarray}}
\newcommand{\eea}{\end{eqnarray}}

\begin{document}

\preprint{IFIC/09-42}

\author{A.~Esteban-Pretel} 
\affiliation{AHEP Group, Institut de F\'{\i}sica Corpuscular -
  C.S.I.C/Universitat de Val\`encia\\
Edifici Instituts d'Investigaci\'o, Apt. 22085, E-46071 Val\`encia, Spain}

\author{R. Tom\`as}
\affiliation{II.\ Institut f\"ur theoretische Physik, Universit\"at
Hamburg, Luruper Chaussee 149, 22761 Hamburg, Germany}

\author{J.~W.~F.~Valle}
\affiliation{AHEP Group, Institut de F\'{\i}sica Corpuscular -
  C.S.I.C/Universitat de Val\`encia\\
Edifici Instituts d'Investigaci\'o, Apt. 22085, E-46071 Val\`encia, Spain}

\title{Interplay between collective effects and non-standard interactions of supernova neutrinos}

\date{\today}

\begin{abstract}
  We consider the effect of non-standard neutrino interactions (NSI,
  for short) on the propagation of neutrinos through the supernova
  (SN) envelope within a three-neutrino framework and taking into
  account the presence of a neutrino background.
  We find that for given NSI parameters, with strength generically
  denoted by $\varepsilon_{ij}$, neutrino evolution exhibits a
  significant time dependence. For $|\varepsilon_{\tau\tau}|\gtrsim$
  $10^{-3}$ the neutrino survival probability may become sensitive to
  the $\theta_{23}$ octant and the sign of $\varepsilon_{\tau\tau}$.
  In particular, if $\varepsilon_{\tau\tau}\gtrsim 10^{-2}$ an
  internal $I$-resonance may arise independently of the matter
  density. For typical values found in SN simulations this takes place
  in the same dense-neutrino region above the neutrinosphere where
  collective effects occur, in particular during the synchronization
  regime. This resonance may lead to an exchange of the neutrino
  fluxes entering the bipolar regime. The main consequences are
    (i) bipolar conversion taking place for normal neutrino mass
    hierarchy and (ii) a transformation of the flux of low-energy
    $\nu_e$, instead of the usual spectral swap.

\end{abstract}


\maketitle

\section{Introduction}
\label{sec:introduction}

Current neutrino data imply that neutrino have mass.  Indeed 
reactor data by the KamLAND collaboration~\cite{Eguchi:2002dm}
and data from accelerator neutrino experiments K2K~\cite{Ahn:2006zz}
and MINOS~\cite{Adamson:2008zt} not only confirm the neutrino flavor
conversion discovered in the study of
solar~\cite{:2008zn,Aharmim:2008kc,abdurashitov:2002nt,Collaboration:2008mr}
and atmospheric neutrinos~\cite{Ashie:2005ik,Ashie:2004mr} but also
indicate that the underlying neutrino flavor conversion mechanism is
oscillatory in both cases~\cite{Pakvasa:2003zv}. An updated review of
the current status of neutrino oscillations is given in
Refs.~\cite{Schwetz:2008er}.

Theories of neutrino mass~\cite{schechter:1980gr,Valle:2006vb}
typically require that neutrinos have non-standard four-Fermi
interactions~\footnote{Other non-standard neutrino properties such as
  electromagnetic transition moments may also be
  present~\cite{Schechter:1981hw}.} which, for short, we denote as
NSI~\cite{Wolfenstein:1977ue,MS,Valle:1987gv}. These are natural
outcome of neutrino mass models and can be of two types:
flavor-changing (FC) and non-universal (NU).
For example, generic seesaw-type models lead to a non-trivial
structure of the lepton mixing matrix characterizing the charged and
neutral current weak interactions implying an effective non-unitary
form for the mixing matrix describing neutrino
oscillations~\cite{schechter:1980gr}.  However, the expected magnitude
of the unitarity violation and of the four-Fermi NSI effects is rather
model-dependent. In the simplest high-scale seesaw models these are
all negligible. In contrast, the 4-Fermi NSI effects induced by the
charged and neutral current gauge interactions may be sizeable in
low-scale seesaw
schemes~\cite{mohapatra:1986bd,Bernabeu:1987gr,Branco:1989bn,Deppisch:2004fa,Malinsky:2005bi,Hirsch:2009mx,Ibanez:2009du,He:2009xd}.

Alternatively, non-standard neutrino interactions may arise in models
where neutrino masses are radiatively induced by low-scale loop
effects~\cite{Zee:1980ai,Babu:1988ki,AristizabalSierra:2007nf}, or
directly ``calculable'' by renormalization group
evolution~\cite{Bazzocchi:2009kc} in some supergravity models.

It is important to realize that the strengths of non-standard
interactions need not be suppressed by the smallness of neutrino
masses. Indeed, they may be very relevant for the propagation of
supernova (SN) neutrinos and give rise to a novel type of resonant
neutrino flavor conversion mechanism that can take place even in the
limit of massless
neutrinos~\cite{Valle:1987gv,Nunokawa:1996tg}~\footnote{Similarly
  flavor and CP-violating effects can survive in the limit of massless
  neutrinos~\cite{Bernabeu:1987gr,Branco:1989bn}.}.
Therefore we argue that, in addition to the precision determination of
the oscillation parameters, it is necessary to test for sub-leading
non-oscillation effects that could arise from non-standard neutrino
interactions in upcoming neutrino
experiments~\cite{Nunokawa:2007qh,Bandyopadhyay:2007kx}. Especially
because NSI effects can in some cases fake genuine mixing
effects~\cite{huber:2001zw,huber:2001de}.

Here we concentrate on the impact of non-standard neutrino
interactions on SN physics. The main motivation of the work is
to reexamine the effect of non-standard neutrino interactions on
supernova when the neutrino self-interaction is taken into
account. 
The extreme conditions under which neutrinos propagate, since they are
created in the SN core until they reach the Earth, may lead to strong
matter effects. It is known that, in particular, the effect of small
values of the NSI parameters can be dramatically enhanced in the inner
strongly deleptonized
regions~\cite{Valle:1987gv,Nunokawa:1996tg,Nunokawa:1996ve,EstebanPretel:2007yu}.
On the other hand, it has been recently pointed
out~\cite{Pastor:2002we, Sawyer:2005jk, Fuller:2005ae, Duan:2005cp,
  Duan:2006an, Hannestad:2006nj, Duan:2007mv, Raffelt:2007yz,
  EstebanPretel:2007ec, Raffelt:2007cb, Raffelt:2007xt, Duan:2007fw,
  Fogli:2007bk, Duan:2007bt, Duan:2007sh, Dasgupta:2008cd,
  EstebanPretel:2007yq, Dasgupta:2007ws, Duan:2008za, Dasgupta:2008my,
  Sawyer:2008zs, Duan:2008eb, Chakraborty:2008zp, Dasgupta:2008cu,
  Fogli:2008fj} that in this region the large neutrino background
could itself affect neutrino propagation significantly.
In this paper we analyze the interplay between these two in principle
coexisting effects, the NSI and the neutrino self-interaction, and the
resulting consequences for neutrino evolution through the SN envelope.

The paper is organized as follows. In Sec.~\ref{sec:eoms} we describe
the equations of motion that govern neutrino propagation taking into
account both NSI and self interaction effects. In
Sec.~\ref{sec:nsi-collective} we study the interplay between these two
effects and analyze the conditions for different regimes of evolution
to take place. In Sec.~\ref{sec:classification} and~\ref{sec:nsi-col},
we describe in detail the salient features characterizing neutrino
evolution in the different regions defined by the conditions given in
the previous section. Finally, before summarizing in
Sec.~\ref{sec:summary}, we give, in Sec.~\ref{sec:discussion}, a time
dependent discussion of the studied effects.

\section{Equations of Motion}                         
\label{sec:eoms}

The equations of motion (EOMs) of the neutrinos traveling through the
SN envelope can be written as 
\begin{equation}
\I\partial_t\varrho_{\bf p}=[{\sf H}_{\bf p},\varrho_{\bf p}]\,,
\label{eq:eoms}
\end{equation}
where $\rho_{\bf p}$ and $\bar\rho_{\bf p}$ represent the density
matrices describing each (anti)neutrino mode.  The diagonal entries
are the usual occupation numbers whereas the off-diagonal terms encode
phase information. The Hamiltonian for neutrinos has the form
\begin{equation}
 {\sf H}_{\bf p}=\Omega_{\bf p}
 +{\sf V}+\sqrt{2}\,G_{\rm F}\!
 \int\!\frac{\D^3{\bf q}}{(2\pi)^3}
 \left(\varrho_{\bf q}-\bar\varrho_{\bf q}\right)
 (1-{\bf v}_{\bf q}\cdot{\bf v}_{\bf p})\,.
\label{eq:hamiltonian}
\end{equation}
For antineutrinos the only difference is $\Omega_{\bf p}\to-\Omega_{\bf p}$.  

\begin{figure}
\begin{center}
\includegraphics[angle=0,width=0.45\textwidth]{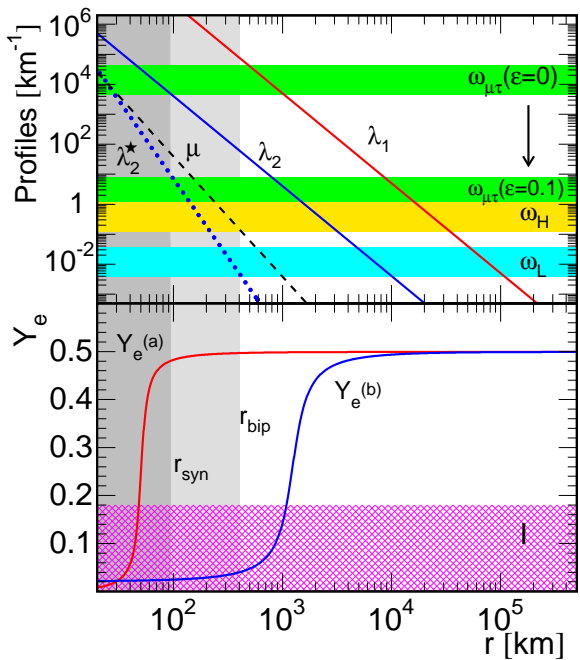}
\caption{Top panel: $\lambda (r)$, profiles as defined in
  Eq.~(\ref{eq:lambda(r)}), for $\lambda_{1,0}= 5\times 10^9$~km$^{-1}$
  ($\lambda_1$) and $\lambda_{2,0}= 4\times 10^6$~km$^{-1}$
  ($\lambda_2$) in solid red and blue lines, respectively;
  $\lambda_2^\star(r)$, as given by Eq.~(\ref{eq:lambda*2}), is shown in
  blue dotted lines; $\mu(r)$, introduced in Eq.~(\ref{eq:mu(r)}), for
  $\mu_0=7\times 10^5~$km$^{-1}$ in black dashed lines; the vacuum
  oscillation frequencies $\omega_{\rm H}$ (yellow band), $\omega_{L}$
  (light blue band), and $\omega^{\rm nsi}_{\mu\tau}$, defined in
  Eq.~(\ref{eq:omega_nsi_mutau}), for $\varepsilon_{\tau\tau}=0$ and 0.1
  (green bands), for energies between 5 and 50 MeV.  The position of
  the synchronization and bipolar radii are also shown. Bottom: radial
  dependence of $Y_e$, given by Eq.~(\ref{eq:Ye}), for two set of parameters:
  $a=0.24,~b=0.165,~r_0=50~(1.2\times 10^3)$~km, and $r_s= 5~(3\times
  10^2)$~km, for $Y_e^{\rm a}~(Y_e^{\rm b})$. The horizontal magenta
  band represents the $Y_e^I$ leading to an internal $I$-resonance for
  $\varepsilon_{\tau\tau}\le 0.1$.  
}
\label{fig:profiles}
\end{center}
\end{figure}

The first term stands for the matrix of vacuum oscillation
frequencies, $\Omega_{\bf p}=U{\rm diag}(m_1^2,m_2^2,m_3^2)/2|{\bf
  p}|U^\dagger$ in the weak basis, where $U$ is the three-neutrino
lepton mixing matrix~\cite{schechter:1980gr} in the unitary
approximation and PDG convention~\cite{Amsler:2008zz}, with no $CP$
phases. We use $\Delta m_{21}^2\equiv m_2^2-m_1^2 = 7.65\times
10^{-5}$~eV$^2$, $|\Delta m_{31}^2|\equiv |m_3^2-m_1^2| = 2.40\times
10^{-3}$~eV$^2$, $\sin^2\theta_{12}=0.3$, as obtained in
e.g. Ref~\cite{Schwetz:2008er}. We consider also
$\sin^2\theta_{13}=10^{-2}$ and three different values for
$\theta_{23}$ in the allowed range at 3$\sigma$, $\sin^2\theta_{23} =$
0.4, 0.5 and 0.6, because our results depend sensitively on
$\theta_{23}$. Given the values of $\Delta m^2$ one can define the two
associated vacuum oscillation frequencies: $\omega_{\rm H}\equiv
\Delta m^2_{31}/2E$ and $\omega_{\rm L}\equiv \Delta m^2_{21}/2E$,
which, in the case of neutrinos with $E=20$~MeV, take the values
$\omega_{\rm H}=0.3$~km$^{-1}$ and $\omega_{\rm L}=0.01$~km$^{-1}$. In
the top panel of Fig.~\ref{fig:profiles} we represent $\omega_{\rm H}$
and $\omega_{L}$ for energies typical in SNe, between 5 MeV and 50
MeV, as yellow and light blue horizontal bands, respectively.

The second term of the Hamiltonian accounts for the interaction of
neutrinos with matter and can be split into two pieces,
\begin{equation}
{\sf V}= {\sf V}_{\rm std} + {\sf V}_{\rm nsi}~.
\end{equation}
The first term, ${\sf V}_{\rm std}$, describes the standard
interaction with matter and can be represented in the weak basis by
${\sf V}_{\rm std}=\sqrt{2}\,G_{\rm F}n_B\,{\rm diag}(Y_e,0,Y_\tau^{\rm eff})$,
with $G_{\rm F}$ the Fermi constant, $n_B$ the baryon density, and
$Y_e=n_e/n_B$ stands for the electron fraction.  We
  consider the following standard weak potential 
\begin{equation}
{\sf V}_{\rm std}=\lambda(r){\rm diag}(Y_e,0,Y_\tau^{\rm eff})\,, 
\label{eq:V}
\end{equation}
with 
\begin{equation}
\label{eq:lambda(r)}
  \lambda(r) =
    \lambda_0\,\left(\frac{R}{r}\right)^3 \,.
\end{equation} 
In the following we assume $R= 10$~km.  In the top panel of
Fig.~\ref{fig:profiles} we show two $\lambda(r)$ profiles for
$\lambda_{1,0} = 5\times 10^9$~km$^{-1}$ and
$\lambda_{2,0}=4\times 10^6$~km$^{-1}$ denoted by $\lambda_1$
and $\lambda_2$, corresponding to typical early and late time
profiles, respectively~\cite{EstebanPretel:2007yq}.

The first element in ${\sf V}^{\rm std}$ represents the charged
current potential and is proportional to the electron fraction,
$Y_e$. According to the SN models, $Y_e$ is characterized by a
transition from a few \% in the inner most deleptonized layers until
values around 0.5 in the outer envelope. Following
Ref.~\cite{EstebanPretel:2007yu} we parametrize it phenomenologically
as
\begin{equation}
Y_e(r) = a + b\arctan[(r-r_0)/r_s]~,
\label{eq:Ye}
\end{equation}
with $a=0.24$ and $b=0.165$. The parameters $r_0$ and $r_s$ describe
where the rise takes place and how steep it is, respectively.
In the bottom
panel of Fig.~\ref{fig:profiles} we show two $Y_e(r)$ profiles for
two different choices of these parameters.
The radius where $\lambda(r)Y_e(r)$ crosses the horizontal bands
$\omega_{\rm H}$ ($\omega_{\rm L}$) determines the well known $H$
($L$) Mikheyev-Smirnov-Wolfenstein (MSW)
resonances~\cite{Mikheev:1986gs,Mikheev:1986wj}. For the $\lambda$ and
$Y_e$ profiles shown in Fig.~\ref{fig:profiles} and energies typical
in SNe the position of both resonances $r_{\rm H}$ and $r_{\rm L}$ lie
above $10^3$~km.

The other non-zero element in ${\sf V}_{\rm std}$ arises from
radiative corrections to neutral-current $\nu_\mu$ and $\nu_\tau$
scattering. Although there are no $\mu$ nor $\tau$ leptons in normal
matter, they appear as virtual states causing a shift $\Delta
V_{\mu\tau}= \sqrt{2}\,G_{\rm F}Y_\tau^{\rm eff}n_B$ between $\nu_\mu$
and~$\nu_\tau$ due to the difference in their masses.  It has the same
effect on neutrino dispersion as real $\tau$ leptons with an effective
abundance~\cite{Botella:1986wy}
\begin{equation}
 Y_\tau^{\rm eff}=\frac{3\sqrt{2}\,G_{\rm F}m_\tau^2}{(2\pi)^2}
 \left[\ln\left(\frac{m_W^2}{m_\tau^2}\right)-1+\frac{Y_n}{3}\right]
 =2.7\times10^{-5}\,,
\label{eq:Ytau}
\end{equation}
where $n_e=n_p$ and $Y_n=0.5$ was assumed. Analogously to the $H$- and
$L$-resonances, the radius where $ \lambda(r)Y_\tau^{\rm eff}\approx
\omega_{\rm H}$ or, equivalently $\lambda(r)\approx\omega_{\rm
  \mu\tau}\equiv \omega_{\rm H}/Y_\tau^{\rm eff}$, defines the
$\mu\tau$-resonance~\cite{Akhmedov:2002zj}.  The upper green band in
the top panel of Fig.~\ref{fig:profiles} represents $\omega_{\mu\tau}$
for energies between 5 and 50 MeV.  In the present analysis we have
not considered the flavor non-universal radiative correction for
neutrino refraction in the presence of a neutrino
background~\cite{Mirizzi:2009td}.

In order to construct the term in the Hamiltonian describing the
non-standard neutrino interactions with a fermion $f$ we parametrize
them with the effective low-energy four-fermion operator:
\begin{equation}
\label{eq:Lnsi}
\mathcal{L}_{NSI} = -\varepsilon^{fP}_{\alpha\beta}
2\sqrt{2}G_F(\bar\nu_\alpha\gamma_\mu L \nu_\beta) (\bar f\gamma^\mu P f)~,
\end{equation}
where $P=L,~R$ and $f$ is a first generation fermion: $e,~u,~d$.  We
neglect the possible NSI effect arising in $\nu-\nu$
interactions~\cite{Blennow:2008er}.  The coefficients
$\varepsilon^{fP}_{\alpha\beta}$ denote the strength of the NSI
between the neutrinos of flavors $\alpha$ and $\beta$ and the
$P-$handed component of the fermion $f$.
Throughout the article we assume neutrinos propagating across
an unpolarized medium~\cite{Nunokawa:1997dp}. Therefore their
evolution in matter will be affected by the vector coupling constant
of the NSI, $\varepsilon_{\alpha\beta}^{fV}=
\varepsilon_{\alpha\beta}^{fL} +
\varepsilon_{\alpha\beta}^{fR}$~\footnote{For the sake of simplicity
  we will omit the superindex $V$.}, see Ref.~\cite{Nunokawa:1997dp}.
We also consider $\varepsilon^f_{\alpha\beta} \in\Re$, neglecting
possible $CP$ violation in the new interactions\footnote{Possible CP
  effects in the case of SUSY radiative corrections on $\mu-\tau$
  neutrino refraction including R-parity breaking interactions have
  been considered in Ref.~\cite{Gava:2009gt}}.
Under these assumptions ${\sf V}_{\rm nsi}$ can be expressed as,
\begin{equation}
({\sf V}_{\rm nsi})_{\alpha\beta} = \sum_{f=e,u,d} ({\sf V}_{\rm
    nsi}^f)_{\alpha\beta}=\sqrt{2}G_FN_f \varepsilon^f_{\alpha\beta} \,,
\end{equation}
where $N_f$ represents the fermion
$f$ number density.
For definiteness and motivated by actual models, for example, those
with broken R parity supersymmetry we take for $f$ the down-type
quark. 
Therefore the NSI potential can be expressed as follows,
\begin{equation}
\label{eq:Vnsi}
({\sf V}_{\rm nsi})_{\alpha\beta} = ({\sf V}_{\rm nsi}^{d})_{\alpha\beta} 
=\varepsilon^{d}_{\alpha\beta}\lambda(r)(2-Y_e)~.
\end{equation}
From now on we will not explicitly write the superindex $d$.
In order to further simplify the problem one can always redefine the
diagonal NSI parameters so that $\varepsilon_{\mu\mu}=0$, since
subtracting a matrix proportional to the identity leaves the physics
involved in the neutrino oscillation unaffected. Thus, in principle at
least five of the six independent $\varepsilon_{\alpha\beta}$
parameters should be taken into account. Nevertheless, the exhaustive
description of NSI in SN neutrinos developed in
Ref.~\cite{EstebanPretel:2007yu} shows that the physics involved can
be described in terms of $\varepsilon'_{\tau\tau}$ and
$\varepsilon'_{e\tau}$, which are just a suitable combination of
$\varepsilon$'s. This motivates us to illustrate the interplay that
could arise between collective effects and NSI by only considering two
non-zero NSI parameters: $\varepsilon_{e\tau}$~and
$\varepsilon_{\tau\tau}$, describing flavor-changing (FC) processes
and non-universality (NU), respectively.
Therefore the term in the Hamiltonian
responsible for the interactions with matter can be written as
\begin{equation}
\label{eq:Vmatrix}
{\sf V}=
\lambda(r)(2-Y_e)
\left(\begin{array}{ccc} \frac{Y_e}{2-Y_e} &
  0 & \varepsilon_{e\tau} \\
0 & 0 & 0 \\
\varepsilon_{e\tau} & 0 & \varepsilon_{\tau\tau}+\frac{Y_\tau^{\rm eff}}{2-Y_e}  
\end{array} \right)~.
\end{equation}
The range of values for the NSI parameters we consider is for the
off-diagonal term $10^{-5}\lesssim |\varepsilon_{e\tau}|\lesssim {\rm
  few}\times 10^{-3}$. This prevents any significant NSI-induced
reduction of the electron fraction $Y_e$ during the core
collapse~\cite{Amanik:2004vm,Amanik:2006ad}.  For the diagonal term we
assume $|\varepsilon_{\tau\tau}|\lesssim 0.1$, allowed by the current
experimental
constraints~\cite{Fornengo:2001pm,Friedland:2004ah,Friedland:2005vy,Escrihuela:2009up}.

Finally, the third term in the Hamiltonian accounts for the collective
flavor transformations induced by neutrino-neutrino interaction.  In
spherical symmetry the EOMs can be expressed as a closed set of
differential equations along the radial
direction~\cite{EstebanPretel:2007ec, Fogli:2007bk}. We solve them
numerically as described in Ref.~\cite{EstebanPretel:2007ec}. The
factor $(1-{\bf v}_{\bf q}\cdot{\bf v}_{\bf p})$ in the Hamiltonian,
${\bf v}_{\bf p}$ being the velocity, implies ``multi-angle effects''
for neutrinos moving on different
trajectories~\cite{Sawyer:2004ai,Sawyer:2005jk,Duan:2006an}. We
consider, though, the single-angle approximation by launching all
neutrinos with $45^\circ$ relative to the radial direction. This
approximation is valid for the neutrino fluxes assumed
  in this analysis provided that the neutrino density exceeds the
  electron density.
The strength of the neutrino-neutrino interaction can be
parametrized by
\begin{equation}
\mu_0 = \sqrt{2}G_{\rm F}(F^R_{\bar\nu_e}-F^R_{\bar\nu_x})\,,
\end{equation}
where the fluxes are taken at the radius~$R$. Following
Ref.~\cite{EstebanPretel:2007ec} we shall assume $\mu_0=7\times
10^5$~km$^{-1}$. In the single-angle approximation the radial
dependence of the neutrino-neutrino interaction strength can be
explicitly written as
\begin{equation}
\label{eq:mu(r)}
\mu(r) = \mu_0 \frac{R^4}{r^4}\frac{1}{2-R^2/r^2}\approx
\mu_0\frac{R^4}{2r^4}\,. 
\end{equation}
In the top panel of Fig.~\ref{fig:profiles} we
show the typical $\mu(r)$ profile for  $\mu_0=7\times
10^5$~km$^{-1}$.
One final property of SN neutrinos with important consequences
  for our study is the hierarchy of fluxes obtained in SN models. 
The typical conditions of the proto-neutron star lead to the
following hierarchy of fluxes $F^R_{\nu_e} > F^R_{\bar\nu_e} >
F^R_{\nu_x}$, with $\nu_x$ standing for
$\nu_\mu,~\nu_\tau,~\bar\nu_\mu$ and $\bar\nu_\tau$.
As in  Ref.~\cite{EstebanPretel:2007ec} we express the lepton
asymmetry with the parameter
$\epsilon=(F^R_{\nu_e}-F^R_{\bar\nu_e})/(F^R_{\bar\nu_e}-F^R_{\bar\nu_x})$. 
Throughout the analysis we shall consider $\epsilon=0.25$.
The equal parts of the fluxes drop out of the
EOMs, so as initial condition we use in the monoenergetic case
$F^R_{\nu_\mu,\bar\nu_\mu,\nu_\tau,\bar\nu_\tau}=0$ and $F^R_{\nu_e}
=(1+\epsilon) F^R_{\bar\nu_e}$.

\section{Non-standard  interactions in the presence of collective effects}
\label{sec:nsi-collective}

In the absence of NSI and collective effects the neutrino propagation
through the SN envelope is basically determined at the well-known MSW
resonances defined in Sec.~\ref{sec:eoms}. They arise when the matter
potential becomes of the same order as the kinetic terms in the
Hamiltonian.  The $L$-resonance occurs always for neutrinos whereas
the $H$-one takes place for (anti)neutrinos for (inverted) normal mass
hierarchy. For our matter profiles and the values of $\theta_{12}$ and
$\theta_{13}$ both resonances are
adiabatic~\cite{Kuo:1989qe,Dighe:1999bi}.  Moreover both involve
electron neutrino flavor and happen in the outer layers of the SN
envelope, see top panel of Fig.~\ref{fig:profiles}.

In addition, the $\mu\tau$-resonance is also
adiabatic~\cite{Akhmedov:2002zj}, but occurs between the $\nu_\mu$
and $\nu_\tau$ or $\bar\nu_\mu$ and $\bar\nu_\tau$ depending on the
neutrino mass hierarchy and the $\theta_{23}$ octant. However, when
considering the neutrino self-interaction this resonance can also
cause significant modifications of the overall $\nu_e$ and $\bar\nu_e$
survival probabilities~\cite{EstebanPretel:2007yq}. According to the
discussion in the previous section the $\mu\tau$-resonance occurs at
\begin{equation}
\label{eq:rmutau}
r_{\mu\tau} \approx R~\left(\frac{\lambda_0 Y_\tau^{\rm eff}}{\omega_{\rm
    H}}\right)^{1/3}= R~\left(\frac{\lambda_0}{\omega_{\rm
    \mu\tau}}\right)^{1/3}\,.
\end{equation}
Due to the smallness of $ Y_\tau^{\rm eff}$ the $\mu\tau$-resonance
happens at deeper layers than the $H$- and $L$-resonances. In
particular, for $\omega_{\rm H}=0.3$~km$^{-1}$, $Y_e=0.5$, and $\lambda_0 =
5\times 10^9$~km$^{-1}$ ($4\times 10^6$~km$^{-1}$)
$r_{\mu\tau}=770$~km (71 km), see the intersection between the upper green
band and the profiles $\lambda_1(r)$ and $\lambda_2(r)$ in the top
panel of Fig.~\ref{fig:profiles}.

The consequence of the addition of an NSI term such as that of
Eq.~(\ref{eq:Vnsi}) is twofold. First, it will affect the MSW
resonances. For the values assumed here the main effect on the $H$-
and $L$-resonances will be just a slight shift in the resonance
point~\cite{Fogli:2002xj,EstebanPretel:2007yu}. The consequences for
the $\mu\tau$-resonance can be more drastic. For sufficiently large
values of $|\varepsilon_{\tau\tau}|$ a negative sign can change the
resonance channel, from $\nu$ to $\bar\nu$ or viceversa, depending on
the octant of $\theta_{23}$. On the other hand, it can significantly
modify the position of the resonance. In the presence of NSI the
$\mu\tau$-resonance happens at
\begin{equation}
r_{\mu\tau} \approx R~\left(\frac{\lambda_0 Y_{\tau,{\rm nsi}}^{\rm
    eff}}{\omega_{\rm H}}\right)^{1/3}=
R~\left(\frac{\lambda_0}{\omega_{\rm \mu\tau}^{\rm nsi}}\right)^{1/3}\,,
\label{eq:rmutaunsi}
\end{equation}
where we have defined
\begin{eqnarray}
Y_{\tau,{\rm nsi}}^{\rm eff} & \equiv & Y_\tau^{\rm
  eff}+(2-Y_e)\varepsilon_{\tau\tau} \,,\\   
\omega^{\rm  nsi}_{\mu\tau} & \equiv &\omega_{\rm H}/Y_{\tau,{\rm nsi}}^{\rm
  eff} \,.
\label{eq:omega_nsi_mutau}
\end{eqnarray}
In particular, for $|\varepsilon_{\tau\tau}|>Y_\tau^{\rm eff}/(2-Y_e)$
the value of $\omega_{\mu\tau}^{\rm nsi}$ decreases, and therefore
$r_{\mu\tau}$ is pushed outwards with respect to the standard case.
The lower green band in the top panel of Fig.~\ref{fig:profiles}
represents $\omega^{\rm nsi}_{\mu\tau}$ for
$\varepsilon_{\tau\tau}=0.1$, $Y_e=0.5$ and typical SN neutrino
energies.  For $E=20$~MeV and the matter profile corresponding to
$\lambda_{2,0}$, the position of the $\mu\tau$-resonance moves out to
a radius of $r_{\mu\tau}\approx 1.3\times 10^3$~km.

The second important consequence is that the new NSI terms can induce
additional resonances~\cite{Valle:1987gv}.
If we consider the inner layers, defined as those where $r\ll r_H$, in
the absence of neutrino self-interaction, one may have $H\approx {\sf
  V}$, as given in Eq.~(\ref{eq:Vmatrix}). One can see that a novel
resonance, which we call $I$-resonance, $I$ standing for ``internal'',
will arise when the condition $H_{ee}=H_{\tau\tau}$ is satisfied.
This occurs when the value of the $\varepsilon_{\alpha\beta}$ is of
the same order as the electron fraction $Y_e$~\cite{Valle:1987gv}.
Current constraints on the $\varepsilon_{\alpha\beta}$'s imply
that small values of $Y_e$ are required for these NSI-induced internal
resonances to occur.  
Hence this condition is only fulfilled in the most deleptonized inner
layers, close to the neutrinosphere, where $Y_e$ reaches values of a
few
\%~\cite{Valle:1987gv,Nunokawa:1996tg,Nunokawa:1996ve,EstebanPretel:2007yu}.
If we neglect the contribution from $Y_\tau^{\rm eff}$ the
corresponding resonance condition can be written as
\begin{equation}
\label{eq:Ye_resI}
Y_e^{I} = \frac{2\varepsilon_{\tau\tau}}{1+\varepsilon_{\tau\tau}}~.
\end{equation}
In the bottom panel of Fig.~\ref{fig:profiles} we show as a horizontal
band the range of $Y_e^{I}$ required for the $I$-resonance to take
place for $\varepsilon_{\tau\tau}\le 0.1$. One sees how for the
$Y_e(r)$ profiles found in numerical simulations the resonance
condition can only be satisfied in the inner layers. For typical
values of $Y_e$ one expects to have the $I$-resonance for
$\varepsilon_{\tau\tau}\gtrsim 10^{-2}$. Moreover, as seen
in~\cite{EstebanPretel:2007yu}, the range of $|\varepsilon _{e\tau}|$
considered ensures adiabaticity.
It must be noted that, in contrast to the standard $H$- and
$L$-resonances, related to the kinetic term, neither the density nor
the energy enter explicitly into the resonance condition, which is
determined only by the electron fraction $Y_e$. Moreover, in contrast
to the standard resonances, the $I$-resonance occurs for both
neutrinos and antineutrinos simultaneously~\cite{Valle:1987gv}.

At the same time, also in the internal region, the neutrino flux
emerging from the supernova core is so dense that, neutrino-neutrino
refraction can cause nonlinear flavor oscillation
phenomena~\cite{Pastor:2002we, Sawyer:2005jk, Fuller:2005ae,
  Duan:2005cp, Duan:2006an, Hannestad:2006nj, Duan:2007mv,
  Raffelt:2007yz, EstebanPretel:2007ec, Raffelt:2007cb,
  Raffelt:2007xt, Duan:2007fw, Fogli:2007bk, Duan:2007bt, Duan:2007sh,
  Dasgupta:2008cd, EstebanPretel:2007yq, Dasgupta:2007ws, Duan:2008za,
  Dasgupta:2008my, Sawyer:2008zs, Duan:2008eb, Chakraborty:2008zp,
  Dasgupta:2008cu}. The crucial effect is a collective mode of pair
transformations of the form $\nu_e\bar\nu_e\to\nu_x\bar\nu_x$, where
$x$ represents some suitable superposition of $\nu_\mu$ and
$\nu_\tau$. This pair-wise form of flavor transformation leaves the
net flavor-lepton number flux unchanged.  For the hierarchy of
neutrino fluxes assumed this conversion occurs only when the neutrino
mass hierarchy is inverted.
Collective flavor transformations start after the synchronization
phase, where $\mu(r_{\rm syn}) \approx 2\omega_{\rm
  H}/(1-\sqrt{1+\epsilon})^2$, and extends a few hundred km in the
so-called bipolar regime until $\mu(r_{\rm bip})\approx \omega_{\rm
  H}$~\cite{Hannestad:2006nj}. At larger radii $\mu(r)<\omega_{\rm H}$
and the neutrino self-interaction becomes negligible. For our chosen
$\mu_0$, an excess $\nu_e$ flux of 25\%, and $\omega_{\rm
  H}=0.3$~km$^{-1}$, we find a synchronization and bipolar radius of
$r_{\rm syn}\simeq 100$~km and $r_{\rm bip}\simeq 330$~km, as
indicated in Fig.~\ref{fig:profiles} by dark and light vertical gray
bands, respectively.
One important consequence of this flavor transformation in the context
of three neutrino flavors is its potential sensitivity to deviations
of $\theta_{23}$ from maximal mixing. As pointed out in
Ref.~\cite{EstebanPretel:2007yq}, in the particular case that the
$\mu\tau$-resonance takes place outside the bipolar radius,
$r_{\mu\tau}\gtrsim r_{\rm bip}$, the final states $\nu_x\bar\nu_x$,
and therefore the final $\nu_e$ and $\bar\nu_e$ survival probability,
depend crucially on the octant of $\theta_{23}$. According to
Eq.~(\ref{eq:rmutaunsi}) and the definition of $r_{\rm bip}$ this
condition is satisfied when
\begin{equation}
\label{eq:rmutaubip}
\lambda_0 Y_{\tau,{\rm nsi}}^{\rm eff} \gtrsim \omega_{\rm H}
\left(\frac{r_{\rm bip}}{R}\right)^3 =
\left[\left(\frac{\mu_0}{2}\right)^3\omega_{\rm H}\right]^{1/4}\sim
1.1\times 10^4~{\rm km}^{-1}\,.
\end{equation}
In the standard case the possibility to discern the $\theta_{23}$
octant only occurs for large density profiles, $\lambda_0\gtrsim
7\times 10^8$~km$^{-1}$, i.e. at early times. This situation would
correspond to the $\lambda_1(r)$ profile in top panel of
Fig.~\ref{fig:profiles}, but not to $\lambda_2(r)$. However the
presence of NSI terms in the Hamiltonian may shift the
$\mu\tau$-resonance to outer layers, making this condition more
flexible. For instance, for $\varepsilon_{\tau\tau}=0.1$ the previous
condition requires only $\lambda_0\gtrsim 1.3\times 10^5$~km$^{-1}$,
see lower green band in the top panel of
Fig.~\ref{fig:profiles}. Therefore the presence of NSI could keep the
possibility to distinguish between the two $\theta_{23}$ octants for
several seconds.


The self-induced flavor transformations however do not occur for
arbitrarily large density profiles. If the electron density $n_e$
significantly exceeds the neutrino density $n_\nu$ in the conversion
region they can be suppressed by matter ~\cite{EstebanPretel:2008ni}.
This is a consequence of neutrinos traveling on different trajectories
when streaming from a source that is not point--like. This multi-angle
matter effect can be neglected if in the collective region, prior
  to  the synchronization radius, we have
\begin{equation}
\label{eq:lambda*}
\lambda^{\star}(r)\equiv Y_e(r)\lambda(r)\frac{R^2}{2r^2} \ll \mu(r)\,.
\end{equation}
The limiting condition can be determined by imposing
  Eq.~(\ref{eq:lambda*}) at $r_{\rm syn}$.  Taking into
account Eqs. (\ref{eq:lambda(r)}) and (\ref{eq:mu(r)}) we obtain
\begin{equation}
\label{eq:lambda*2}
Y_e(r_{\rm syn})\lambda_0\frac{R}{r_{\rm syn}} \ll \mu_0\,.
\end{equation}
If we assume $Y_e=0.5$ and $r_{\rm syn}=100$~km then this condition
amounts to $\lambda_0\ll 1.4\times 10^7$~km$^{-1}$. 
In the examples considered in the top panel of Fig.~\ref{fig:profiles}
this is fulfilled for $\lambda_{2,0}$.  The condition
$\lambda_2^\star(r)\ll \mu(r)$ is then satisfied in the bipolar
region. As a consequence collective effects are not matter
suppressed. This is not the case of $\lambda_1(r)$.
In the standard case the limiting $\lambda_0=1.4\times
10^7$~km$^{-1}$, above which multi-angle matter effects suppress the
collective effects, is, though, smaller than the minimum
$\lambda_0=7\times 10^8$~km$^{-1}$ required for the $\mu\tau$ effect
to be important.  The situation could drastically change in the
presence of NSI. As previously discussed, non-zero NSI non-universal
(diagonal) parameters could help moving the $\mu\tau$ resonance out of
the $r_{\rm syn}$ even for $\lambda_0$ smaller than $1.4\times
10^7$~km$^{-1}$. The consequence is that large enough
$|\varepsilon_{\tau\tau}|$ would make the neutrino propagation through
the SN envelope highly sensitive to the $\theta_{23}$ octant.

\section{Classification of regimes}
\label{sec:classification}

\begin{figure}[h!]
\begin{center}
\includegraphics[angle=0,width=0.45\textwidth]{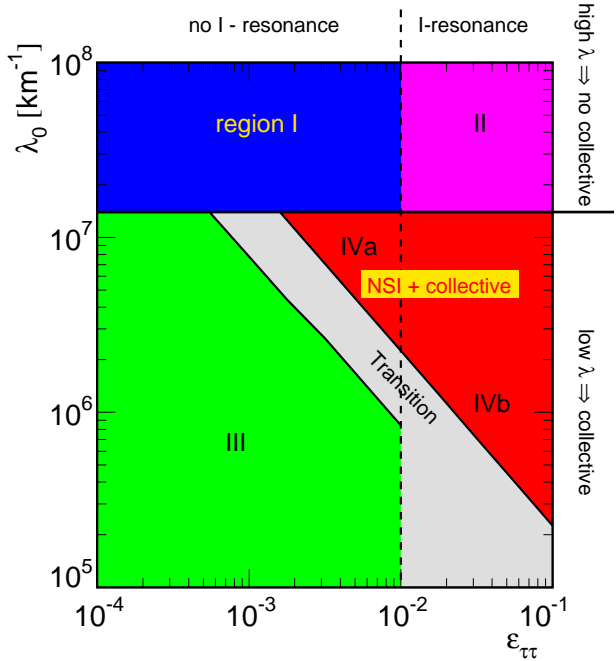}
\caption{Different regimes of the neutrino propagation depending on
  the value of $\lambda_0$ and $\varepsilon_{\tau\tau}$, as described
  in the text.
}
\label{fig:regions}
\end{center}
\end{figure}

In this section we summarize all the information formerly
introduced. Taking into account the conditions given in
Eqs.~(\ref{eq:Ye_resI}), (\ref{eq:rmutaubip}), (\ref{eq:lambda*}), and
(\ref{eq:lambda*2}) we can roughly identify four different regimes of
the neutrino propagation depending on $\lambda_0$ and
$\varepsilon_{\tau\tau}$. This scheme is displayed in
Fig.~\ref{fig:regions} for $\epsilon=0.25$ and $\mu_0=7\times
10^5$~km$^{-1}$. It is important to notice that the size of the
different regions depends on the particular values of 2$\epsilon$
and $\mu_0$. Nevertheless, unless kinematic decoherence between
different angular modes is triggered~\cite{EstebanPretel:2007ec}, this
schematic picture certainly holds.

To first approximation the four regions can be defined in terms of
matter suppression (or not) of collective effects and presence (or
not) of the internal $I$-resonance.  Equation~(\ref{eq:lambda*2}) is
depicted as a horizontal solid line at $\lambda_0=1.4\times
10^7$~km$^{-1}$. For higher $\lambda_0$ matter suppresses collective
effects whereas for smaller densities collective effects are
present. For intermediate values, $\lambda^\star(r)\sim\mu(r)$, there
would be a matter induced decoherence~\cite{EstebanPretel:2008ni}. To
make the discussion as simple as possible we will only consider the
extreme cases.
On the other hand the vertical dashed line at
$\varepsilon_{\tau\tau}=10^{-2}$ indicates the presence (right) or
absence (left) of the NSI-induced $I$-resonance.
This simple scheme becomes further complicated if one adds the
possibility that the $\mu\tau$-resonance lies outside the bipolar
region. In the next subsections we analyze in detail the different
possibilities.

\subsection{Region I}
\label{subsec:regionI}

On the upper left corner we have the region I, defined by $\lambda_0
\gtrsim 1.4\times 10^7$~km$^{-1}$ and $\varepsilon_{\tau\tau}\lesssim
10^{-2}$. According to the previous discussion, this range of
parameters leads to no collective effects, since they are suppressed
by matter, and no $I$-resonance. Assuming that the $L$- and
$H$-resonances are adiabatic the $\nu_e$ and $\bar\nu_e$ survival
probability is then only fixed by the mass hierarchy.  The NSI terms
will lead at most to a small shift in its
position~\cite{Fogli:2002xj,EstebanPretel:2007yu}.

\begin{figure}
\begin{center}
\includegraphics[angle=0,width=0.45\textwidth]{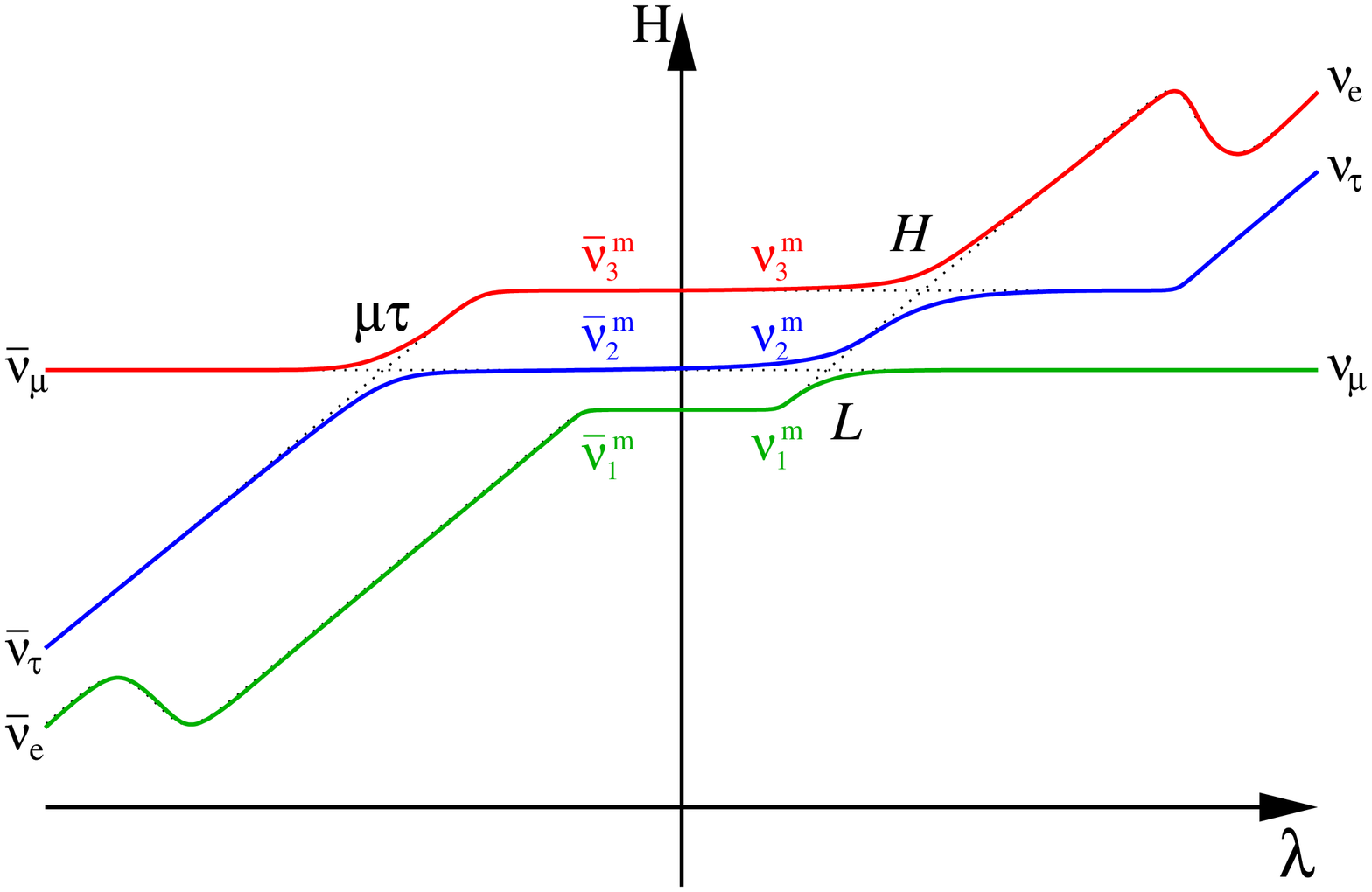}
\vskip6pt
\includegraphics[angle=0,width=0.45\textwidth]{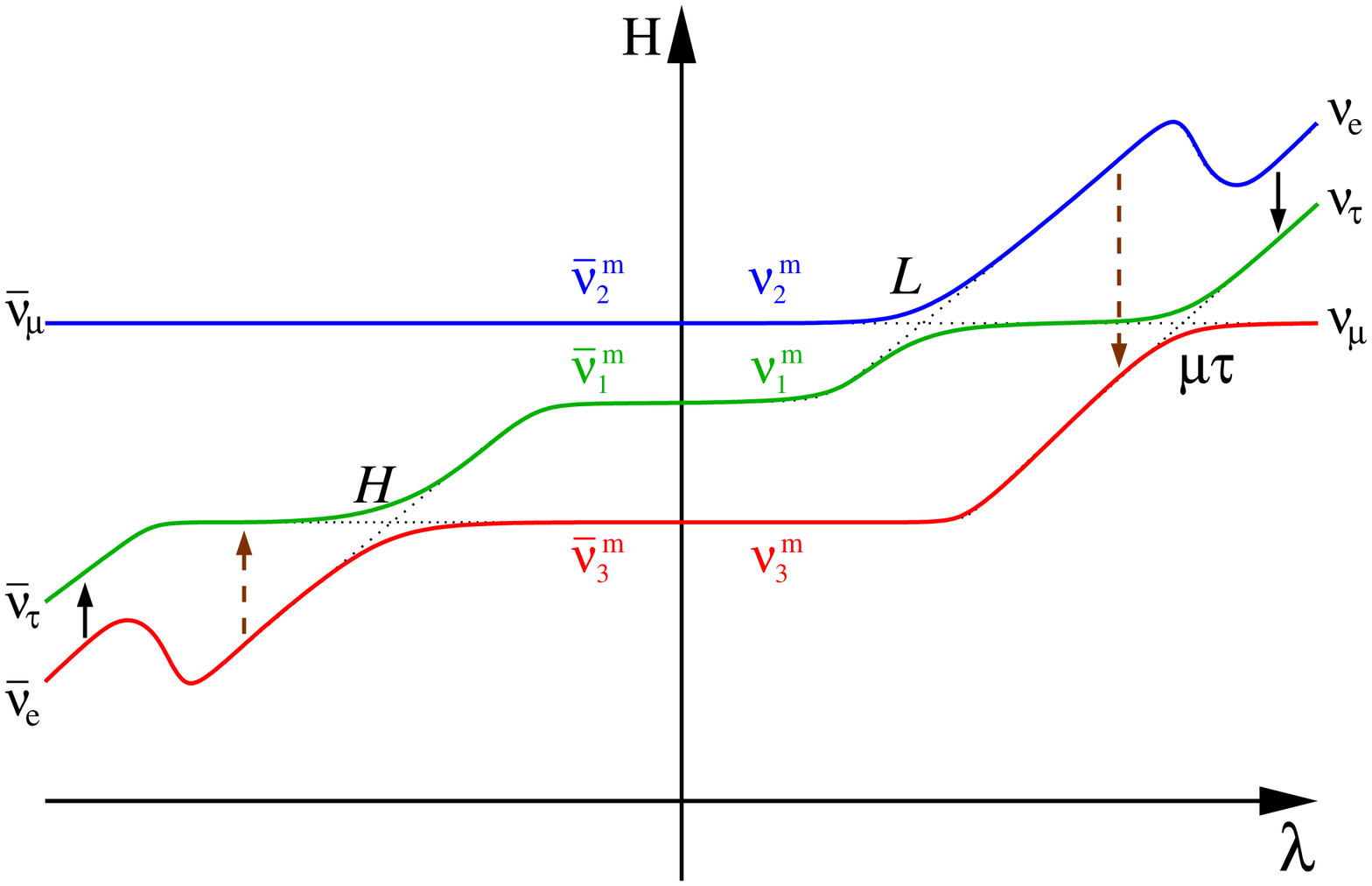}
\vskip6pt
\includegraphics[angle=0,width=0.45\textwidth]{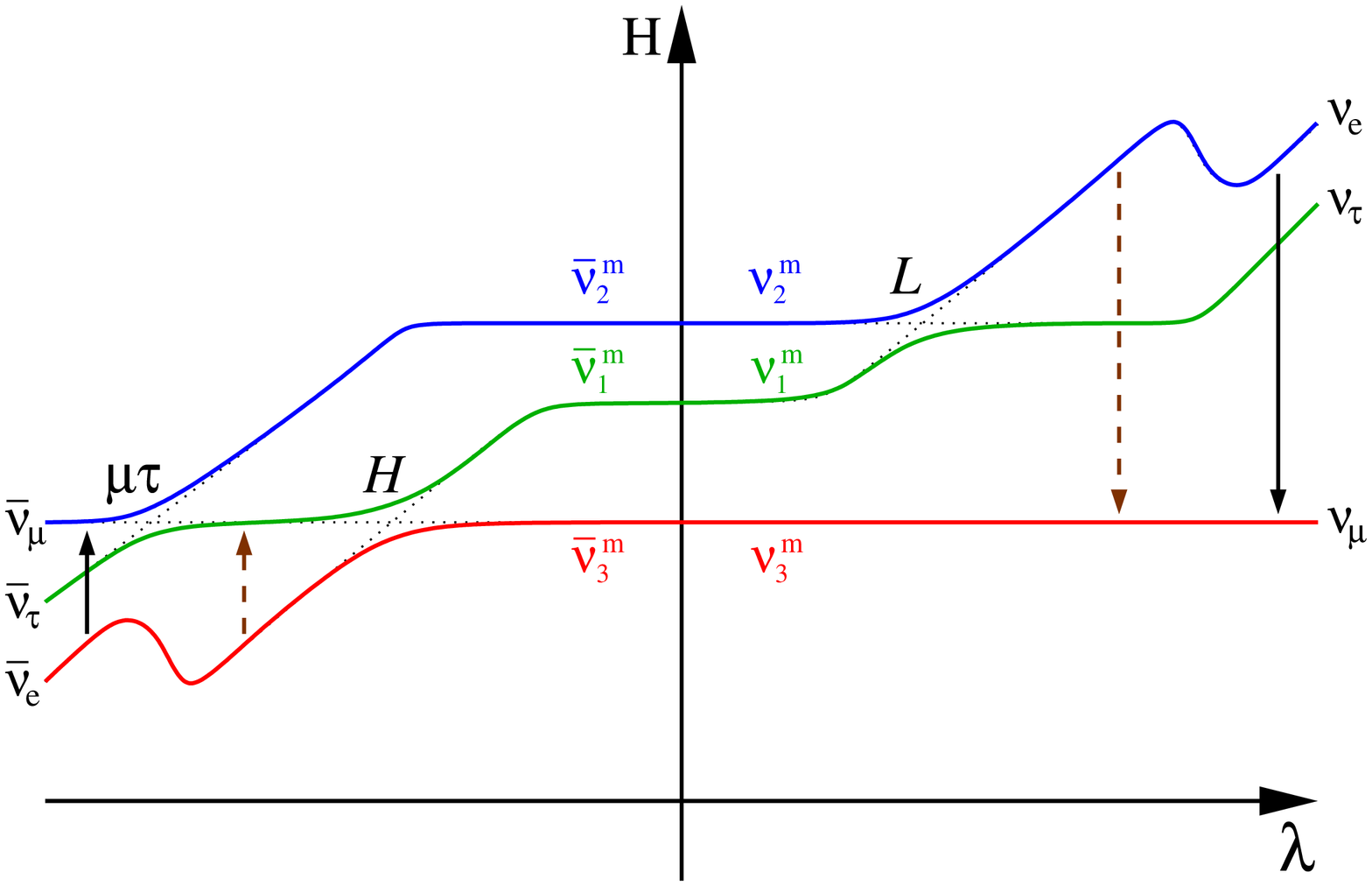}
\caption{ Level crossings in the absence of $I$-resonance for normal
  and $\theta_{23} <\pi/4$
  (top), and inverted mass hierarchy for $\theta_{23} <\pi/4$
  (middle), and $\theta_{23} > \pi/4$ (bottom). The dashed and
  solid arrows in the middle and bottom indicate the pair
  transformations due to collective effects happening after (dashed)
  or before (solid) the $\mu\tau$-resonance.  }
\label{fig:levcros_noI}
\end{center}
\end{figure}

\begin{figure}
\begin{center}
\includegraphics[angle=0,width=0.45\textwidth]{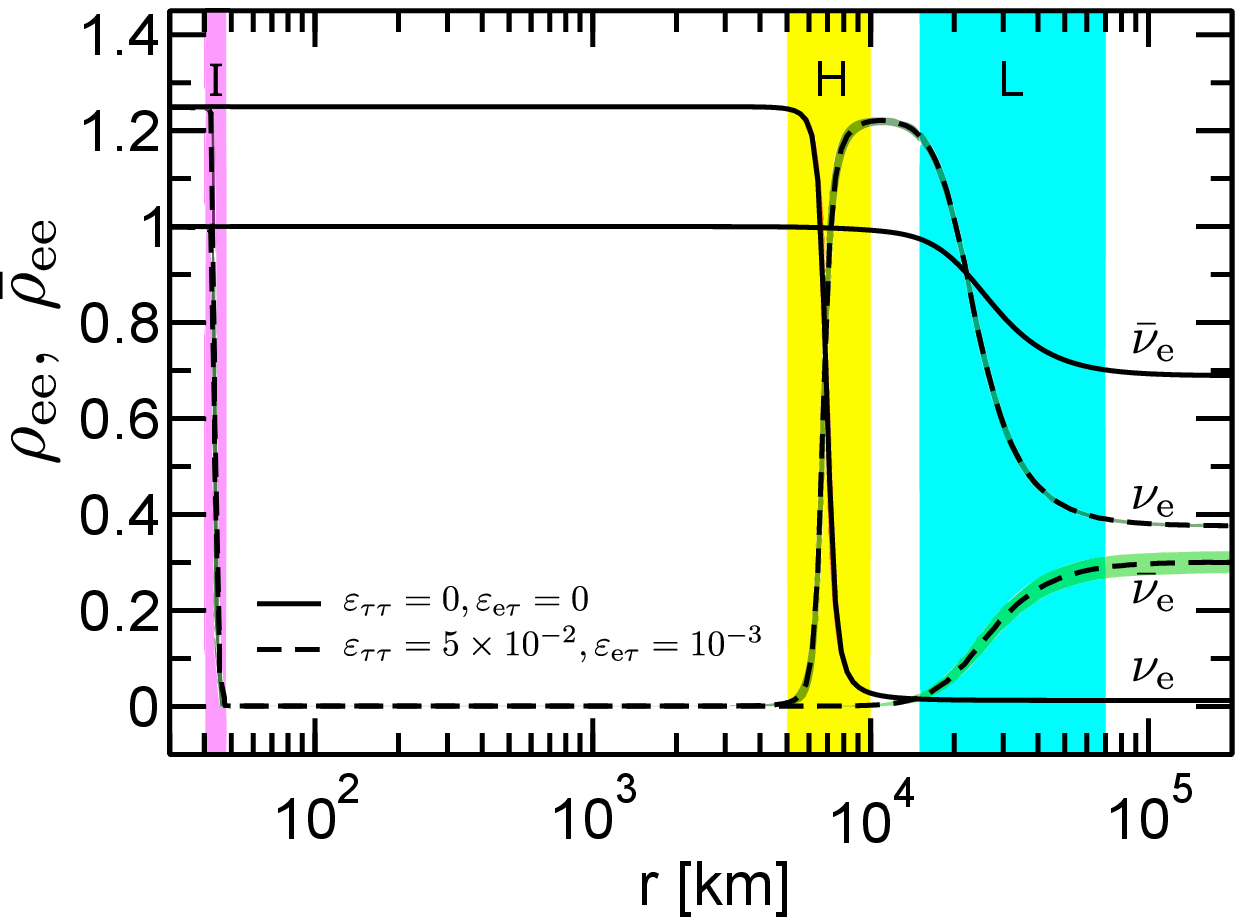}
\vskip8pt
\includegraphics[angle=0,width=0.45\textwidth]{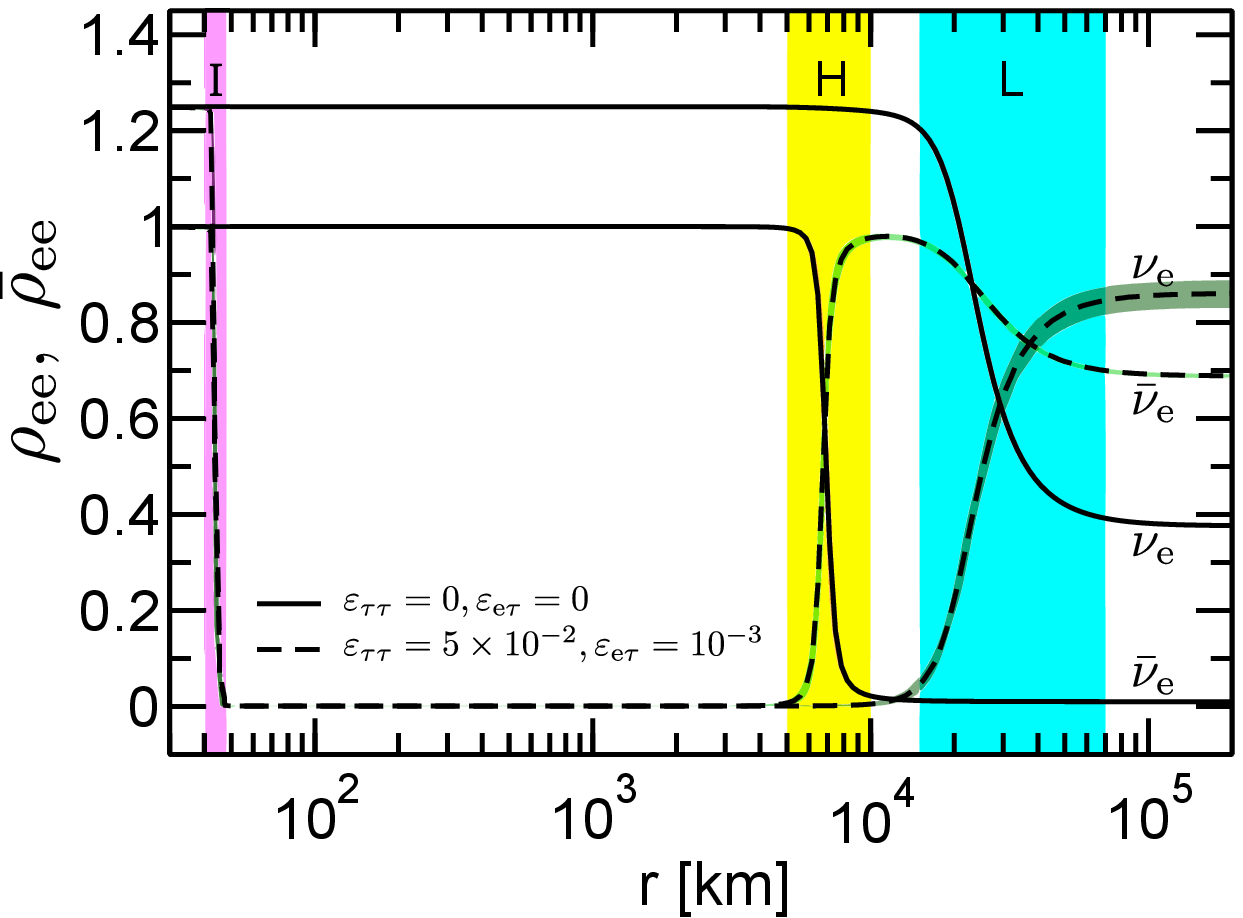}
\caption{Radial dependence of density matrix elements $\rho_{ee}$ and
  $\bar\rho_{ee}$ corresponding to regions I (solid) and II (dashed)
  in Fig.~\ref{fig:regions}.  Top panel represents normal mass
  hierarchy and bottom panel inverted mass hierarchy. We assume
  $\lambda_0=1\times 10^8$~km$^{-1}$, $\omega_{\rm H}=0.3$~km$^{-1}$,
  and $\sin^2\theta_{23}=0.5$. Vertical bands indicate regions where
  resonances take place. }
\label{fig:rho_regionI-II}
\end{center}
\end{figure}

In Fig.~\ref{fig:levcros_noI} we show the level crossing scheme for
normal (top), and inverted mass hierarchy for $\theta_{23} < \pi/4 $
(middle), and $\theta_{23} > \pi/4$ (bottom). It is
  remarkable to note the presence of a dip in the uppermost curves at high
  densities, corresponding to the $V_{ee}$ entry in
  Eq.~(\ref{eq:Vmatrix}). This is a direct consequence of the radial
  dependence of $Y_e(r)$, as given in Eq.~(\ref{eq:Ye}).  The arrows
in these figures represent the transitions caused by the collective
effects, and must therefore be ignored when these are not present. In
the normal hierarchy case $\nu_e$ and $\bar\nu_e$ leave the SN as
$\nu_3$ and $\bar\nu_1$, whereas for inverted mass hierarchy they
escape as $\nu_2$ and $\bar\nu_3$ for any octant. The survival
probabilities can then be written as $P(\nu_e\rightarrow\nu_e)\approx
\sin^2\theta_{13}~(\sin^2\theta_{12})$ and
$P(\bar\nu_e\rightarrow\bar\nu_e)=
\cos^2\theta_{12}~(\sin^2\theta_{13})$ for normal (inverted) mass
hierarchy.
Figure~\ref{fig:rho_regionI-II} represents in solid lines the radial
evolution of $\rho_{ee}$ and $\bar\rho_{ee}$ assuming
$\lambda_0=1\times 10^8$~km$^{-1}$, $\omega_{\rm H}=0.3$~km$^{-1}$,
$\sin^2\theta_{23}=0.5$, and $\epsilon=0.25$. The vertical bands
indicate where the resonance conversions take place. In order to
perform the plot we have artificially set $\mu_0=0$.  We want to
remind here that both $\rho_{ee}$ and $\bar\rho_{ee}$ are normalized
to the $\bar\nu_e$ flux, and therefore, while $\bar\rho_{ee}$
corresponds directly to $\bar\nu_e$ survival probability, $\rho_{ee}$
must be corrected by a factor $(1+\epsilon)$ in order to obtain the
corresponding survival probability,
$\rho_{ee}=P(\nu_e\rightarrow\nu_e)(1+\epsilon)$.


\subsection{Region II}
\label{subsec:regionII}

\begin{figure}
\begin{center}
\includegraphics[angle=0,width=0.45\textwidth]{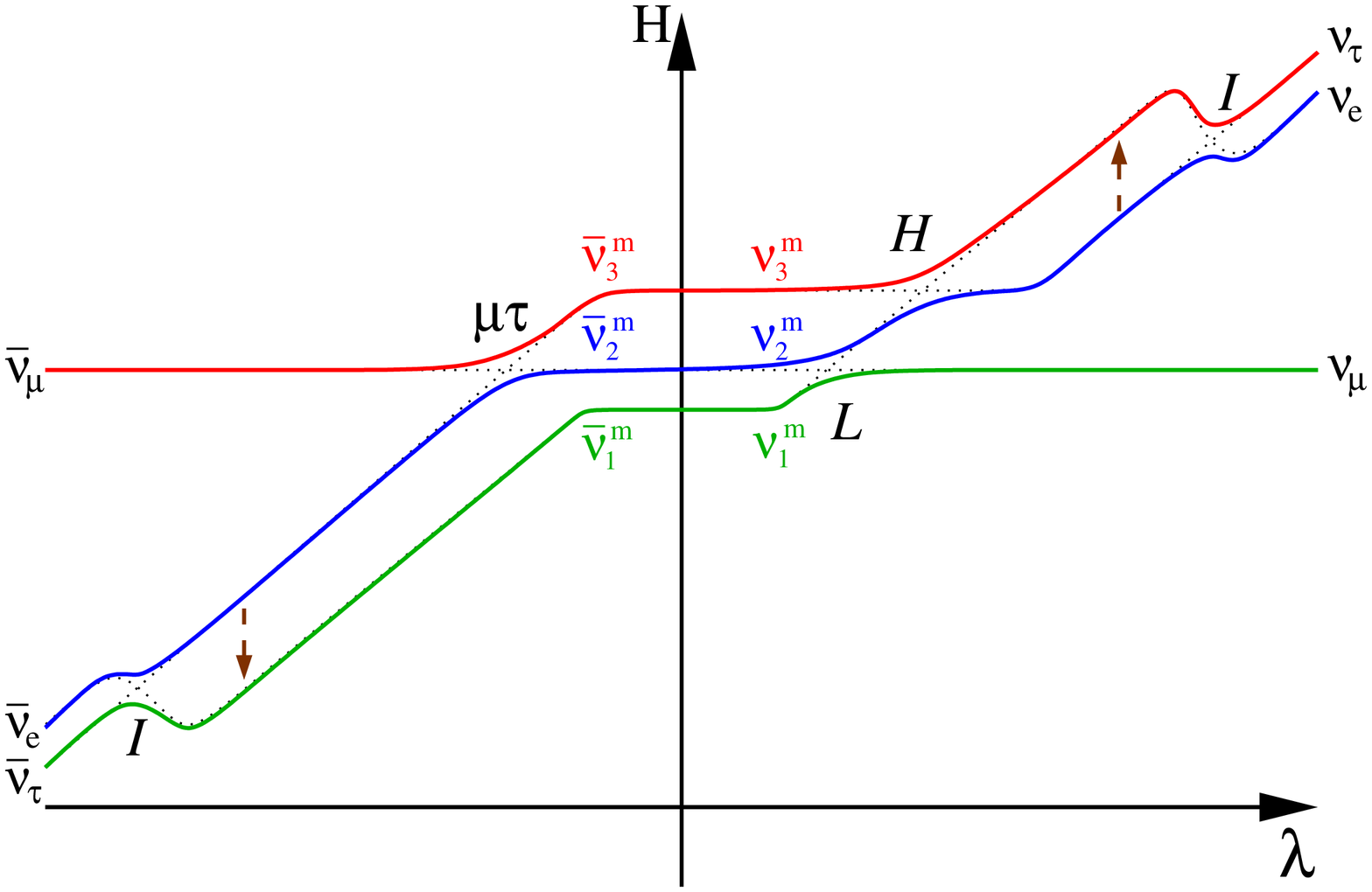}
\vskip6pt
\includegraphics[angle=0,width=0.45\textwidth]{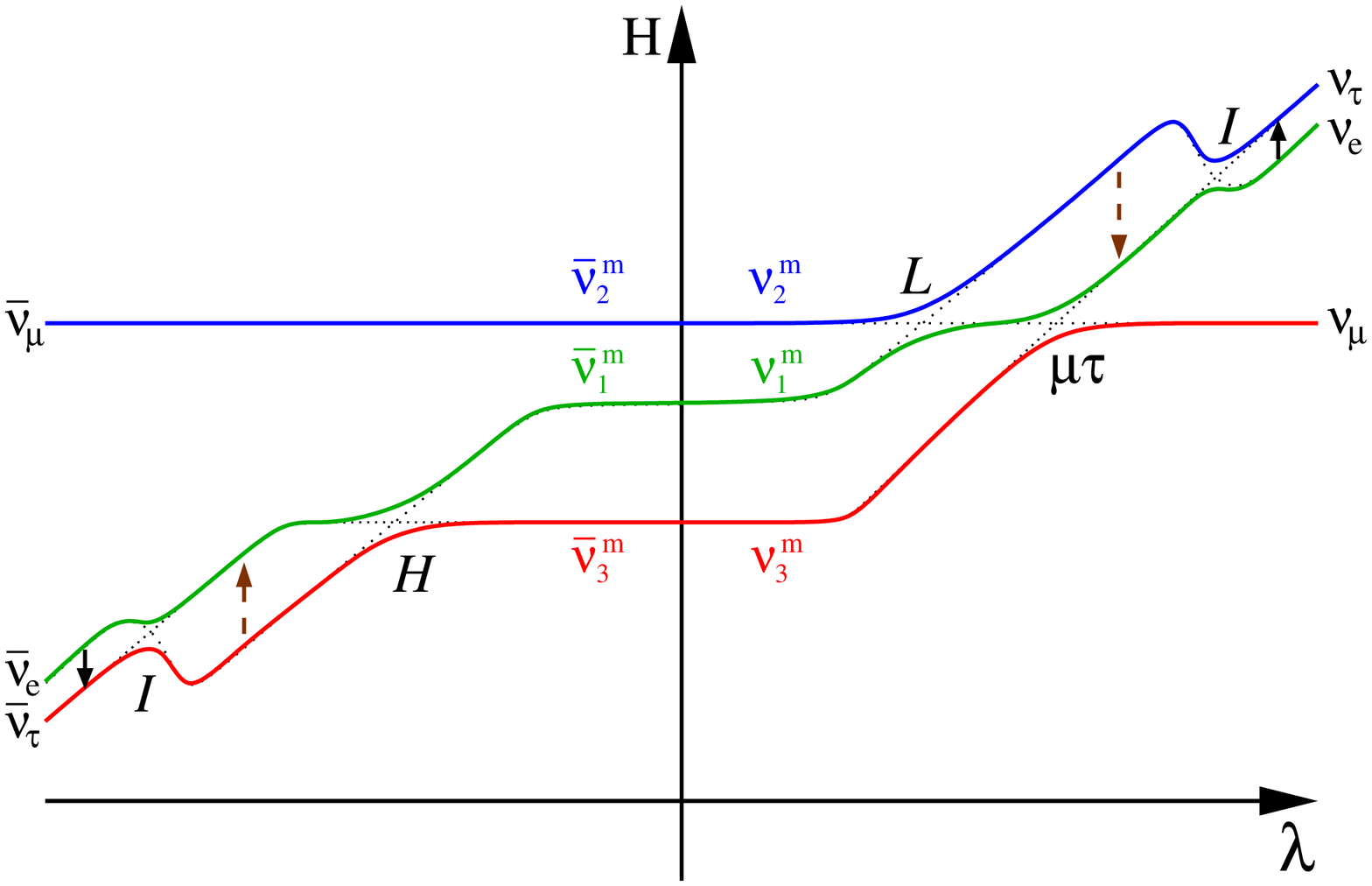}
\vskip6pt
\includegraphics[angle=0,width=0.45\textwidth]{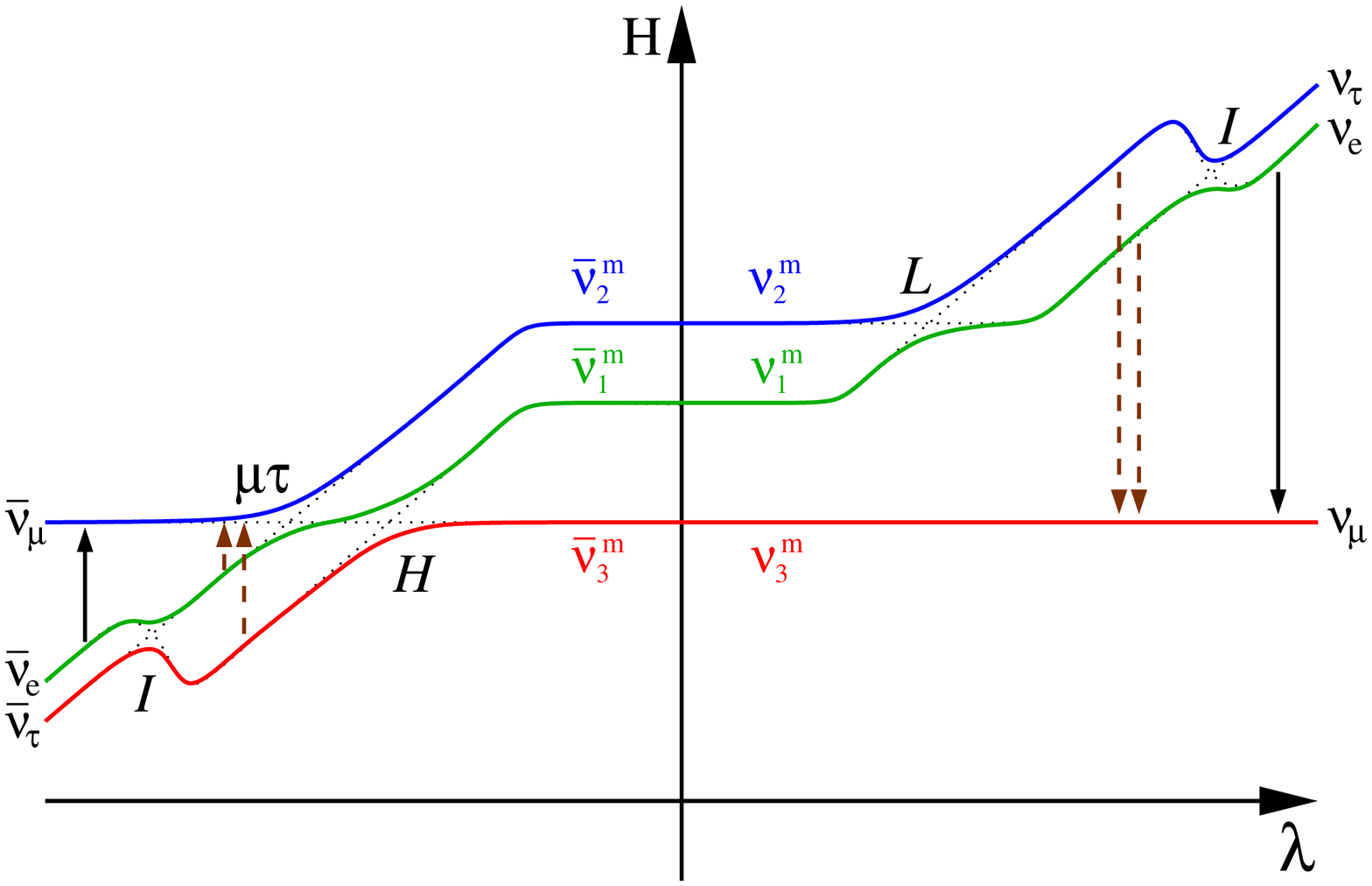}
\caption{Same as Fig.~\ref{fig:levcros_noI} in the presence of
  $I$-resonance. Dashed arrows represent pair transformations due to
  collective effects occurring after the $I$-resonance. In the bottom
  panel there are two possibilities for this transition, corresponding
  to an adiabatic or non-adiabatic $I$-resonance.  The solid arrows in
  the middle and bottom panel refer to collective effects happening
  before the $I$-resonance.}
\label{fig:levcros_I}
\end{center}
\end{figure}

The region II, on the upper right corner, is defined by $\lambda_0
\gtrsim 1.4\times 10^7$~km$^{-1}$ and $\varepsilon_{\tau\tau}\gtrsim
10^{-2}$. As in region I the matter density is so high that prevents
neutrinos from undergoing collective effects. However, the values of
the diagonal NSI terms in this region are large enough to fulfill
Eq.~(\ref{eq:Ye_resI}), causing the $I$-resonance to appear.
In contrast to the previous case $\nu_e$ and $\bar\nu_e$ are now
created as $\nu_2^{\rm m}$ ($\nu_1^{\rm m}$) and $\bar\nu_2^{\rm m}$
($\bar\nu_1^{\rm m}$) for normal (inverted) mass
hierarchy
\footnote{The different neutrino basis used
    in the analysis are: $\nu_\alpha$ with $\alpha=e,~\mu,~\tau$
    representing the flavor basis; $\nu_i$ with $i=1,~2,~3$ standing
    for the mass basis, diagonalizing the Hamiltonian in vacuum,
    $H=\Omega$;  the matter basis, $\nu^{\rm m}_i$, diagonalizing the
    Hamiltonian in absence of neutrino self-interactions $H=\Omega +
    {\rm V}$; and the basis $\nu_i'$ defined as the one diagonalizing
the Hamiltonian in vacuum in the case that all mixing angles are
very small ({\em after rotating the matter term away}).}.
They cross
adiabatically all resonances and leave the SN as $\nu_2$ ($\nu_1$) and
$\bar\nu_2$ ($\bar\nu_1$) for normal (inverted) mass
hierarchy~\cite{EstebanPretel:2007yu}.  In Fig.~\ref{fig:levcros_I} we
show the level crossing scheme analogous to Fig.~\ref{fig:levcros_noI}
but in the presence of the $I$-resonance. The survival probabilities
are now
$P(\nu_e\rightarrow\nu_e)=P(\bar\nu_e\rightarrow\bar\nu_e)\approx
\sin^2\theta_{12}~(\cos^2\theta_{12})$ for normal (inverted) mass
hierarchy.
The black dashed lines in Fig.~\ref{fig:rho_regionI-II} show the
expected radial evolution of $\rho_{ee}$ and $\bar\rho_{ee}$,
respectively, when neutrinos and antineutrinos undergo an adiabatic
$I$-resonance. The width of the green band accounts for the
oscillatory behavior of neutrinos.
As for region I, we have made the calculation assuming
$\mu_0=0$. However, we have analyzed the single energy and multi-angle
case within two-flavor framework for the range of parameters here
discussed, and have verified that collective effects are indeed
suppressed and the $I$-resonance is present for both normal and
inverted hierarchies. This means that the behavior in region II
corresponds indeed to the case discussed in
Ref.~\cite{EstebanPretel:2007yu}.

\subsection{Region III}
\label{subsec:regionIII}

Let us now consider the lower part of Fig.~\ref{fig:regions},
i.e. when $\lambda_0\lesssim 1.4\times 10^7$~km$^{-1}$. The main
feature of this scenario is the presence of collective effects. As
discussed in Ref.~\cite{EstebanPretel:2007yq}, and here reviewed,
these in turn depend on the relative position of the
$\mu\tau$-resonance with respect to the synchronization and bipolar
radius. We can then distinguish two different regimes:
On the bottom left corner we define region III by the condition
$r_{\mu\tau}\lesssim r_{\rm bip}$, and on the bottom right corner we
have region IV defined by $r_{\mu\tau}\gtrsim r_{\rm bip}$. In the
middle of both there is a transition region whose width we have
determined numerically.

\begin{figure}
\begin{center}
\includegraphics[angle=0,width=0.45\textwidth]{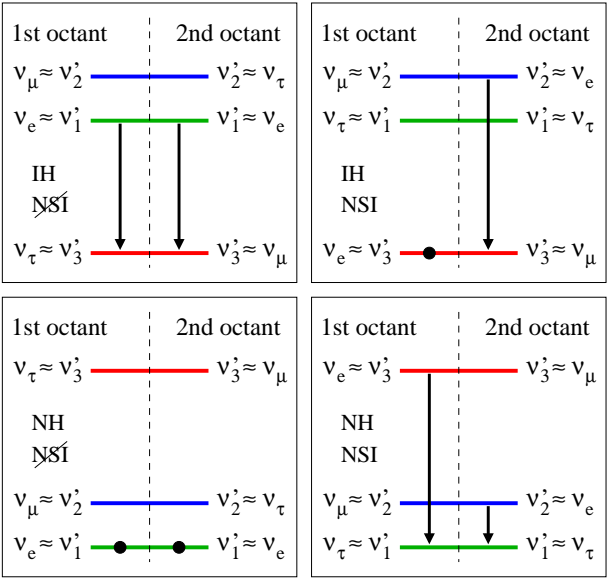}
\caption{Neutrino vacuum level diagrams without NSI
    effects (left panels) and with the NSI-induced $I$ resonance
    occurring before the bipolar conversion (right panels). The top
    (bottom) panels correspond to inverted (normal) neutrino mass
    hierarchy. In each panel the two possible $\theta_{23}$ are
    indicated. The 12 and 13-mixing angles are assumed to be very
    small, mimicking the effect of ordinary matter. The effect of
    collective conversions is indicated by an arrow.}
\label{fig:vaclevels}
\end{center}
\end{figure}

\begin{figure}
\begin{center}
\includegraphics[angle=0,width=0.45\textwidth]{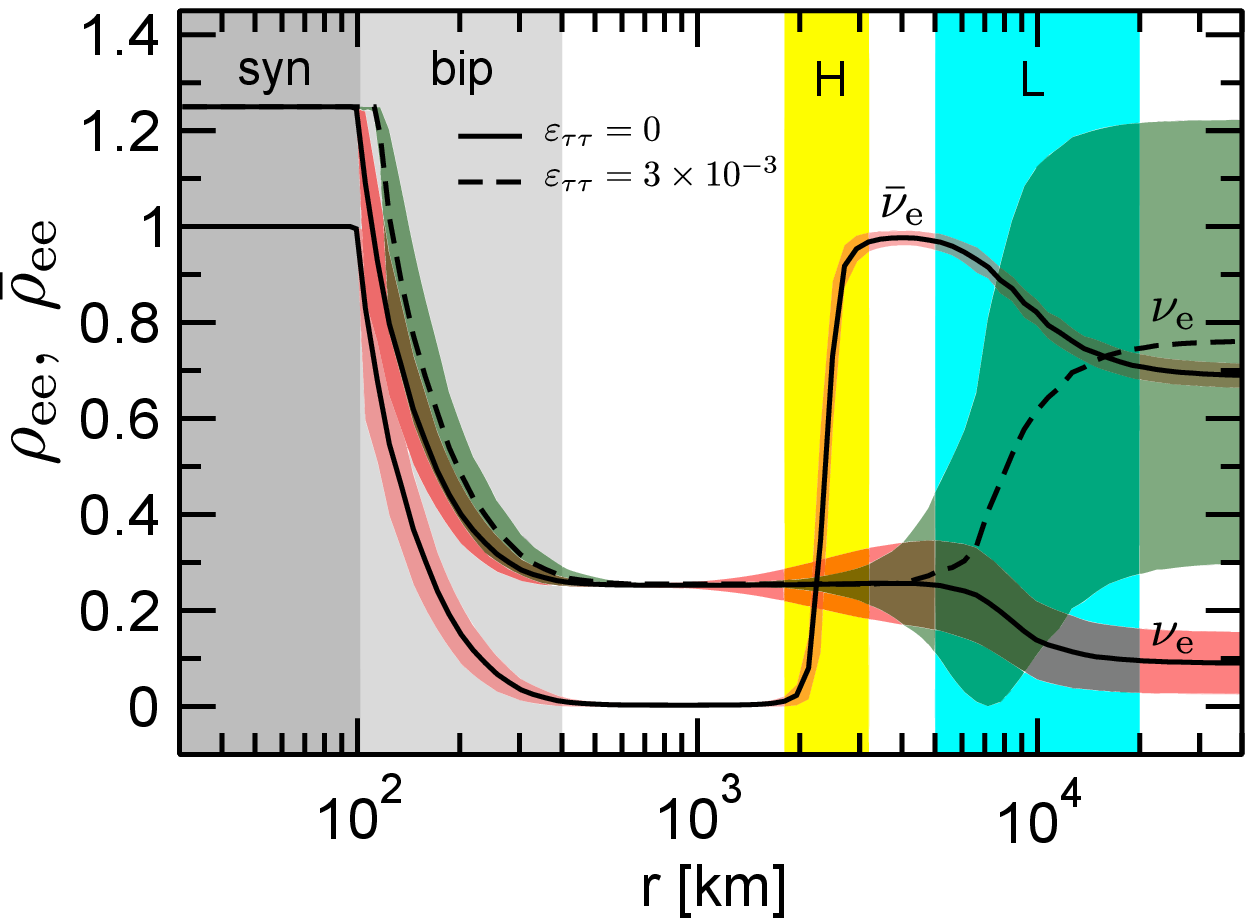}
\vskip8pt
\includegraphics[angle=0,width=0.45\textwidth]{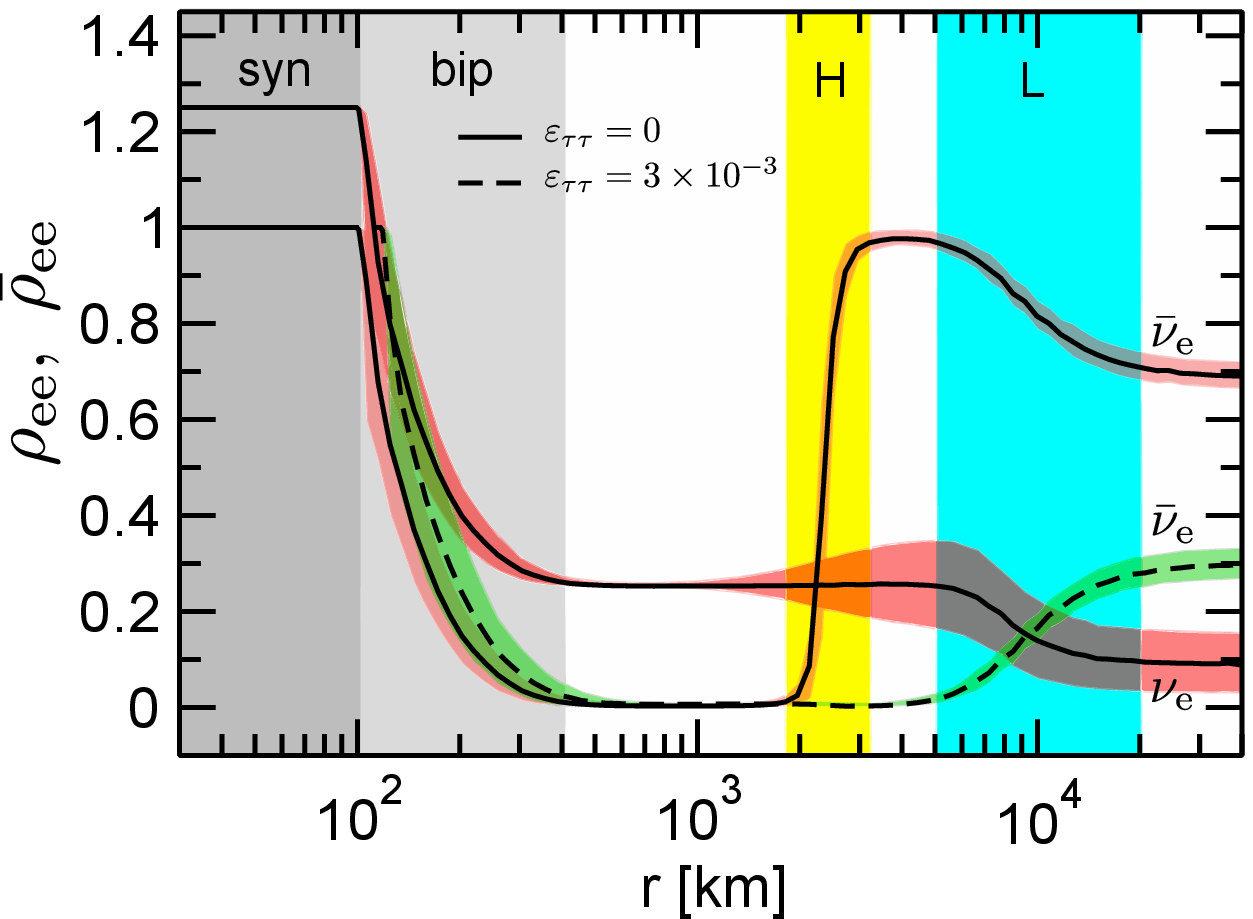}
\caption{Radial dependence of $\rho_{ee}$ and $\bar\rho_{ee}$ for
  region III with $\varepsilon_{\tau\tau}=0$ (solid) and IVa (dashed)
  with $\varepsilon_{\tau\tau}= 3\times 10^{-3} $ for inverted mass
  hierarchy, and $\sin^2\theta_{23}=0.4~(0.6)$ in the top (bottom
  panel). In both cases $\lambda_0=4\times 10^6$~km$^{-1}$ and
  $\varepsilon_{e\tau}=0$. The bands around the lines represent
  modulations. Vertical gray bands stand for synchronized (dark) and
  bipolar (light) regime. Resonance regions are also displayed.}
\label{fig:rho_regionIII-IVa}
\end{center}
\end{figure}

Let us now discuss region III. According to Eq.~(\ref{eq:rmutaubip})
this range of parameters satisfies the condition
\begin{equation}
\lambda_0[Y_\tau^{\rm eff} + (2-Y_e)\varepsilon_{\tau\tau}]\lesssim
1.1\times 10^4~{\rm km}^{-1}\,,
\end{equation}
which, for $Y_e=0.5$, roughly amounts to
$\lambda_0\varepsilon_{\tau\tau}\lesssim 7.3\times 10^3$~km$^{-1}$,
see Fig.~\ref{fig:regions}.  This situation can be reduced to the
standard two-flavor scenario. In order to better understand the
consequences of collective effects it is convenient to realize that
the impact of ordinary matter can be transformed away by going into a
rotating reference frame for the polarization
vectors~\cite{Duan:2005cp, Hannestad:2006nj}. Collective conversions
proceed in the same way as they would in the absence of the matter
term $V_{\rm std}$, except that the effective mixing angle is reduced.
After this rotation the mass eigenstates $\nu_i'$ now
  approximately coincide with the interaction eigenstates
  $\nu_\alpha$. In particular, 
the initial states $\nu_e$ and $\bar\nu_e$ can therefore be
essentially identified with $\nu_1'$ and $\bar\nu_1'$,
respectively.  If the neutrino mass hierarchy is normal, we begin in
the lowest-lying state and no collective effects take
place, see bottom left panel of
  Fig.~\ref{fig:vaclevels} (also in Fig. 4 of
  Ref.~\cite{EstebanPretel:2007yq}).  The situation is then similar to
that in region I, see top panel of Fig.~\ref{fig:rho_regionI-II}.
However in the case of inverted mass hierarchy both $\nu_1'$ and
$\bar\nu_1'$ correspond to the intermediate state. The effect of the
self-interaction is to drive them to the lowest-lying states, which in
this case are $\nu_3'$ and $\bar\nu_3'$, see top left
  panel of Fig.~\ref{fig:vaclevels}. In terms of
    matter eigenstates $\nu_i^{\rm m}$ this is 
shown by dashed arrows in the middle and bottom panels of
Fig.~\ref{fig:levcros_noI}.
In the case of $\nu_e$ a fraction equal to $\epsilon F_{\bar\nu_e}$ is
not transformed and stays in $\nu_2^{\rm m}$ and evolves as in the
absence of neutrino-neutrino interactions, i.e. adiabatically through
the $L$-resonance. The rest of $\nu_e$ are transformed to $\nu_3^{\rm
  m}$. As a result the final $\nu_e$ flux, normalized to the initial
$\bar\nu_e$ flux, is expected to be approximately $\rho^{\rm
  final}_{ee}=\epsilon \sin^2\theta_{12}+\sin^2\theta_{13}\simeq
0.08$. On the other hand, after the pair transformation $\bar\nu_e$
cross the $H$-resonance adiabatically and leave the star as $\nu_1$,
leading to a final normalized flux of approximately $\bar\rho^{\rm
  final}_{ee}=\cos^2\theta_{12}\simeq 0.7$.
This can be seen in Fig.~\ref{fig:rho_regionIII-IVa}, where we show in
solid lines the radial evolution of $\nu_e$ and $\bar\nu_e$ for
inverted mass hierarchy assuming
$\lambda_0=4\times 10^6$~km$^{-1}$, $\omega_{\rm H}=0.3$~km$^{-1}$,
and $\sin^2\theta_{23}=0.4$ (top) and 0.6 (bottom). 
As can be seen in the figure, the result is independent of the
$\theta_{23}$ octant.

\subsection{Region IV}
\label{subsec:regionIV}

Finally, neutrinos with parameters in the right bottom corner (region
IV) will feel both collective and NSI effects. This region of
parameters is defined by the condition that the $\mu\tau$-resonance
lies outside the bipolar region. According to Eq.~(\ref{eq:rmutaubip})
this amounts to 
\begin{equation}
\lambda_0[Y_\tau^{\rm eff} + (2-Y_e)\varepsilon_{\tau\tau}]\gtrsim
1.1~\times 10^4~{\rm km}^{-1}\,.
\end{equation}
As discussed above, for the standard case and $Y_e=0.5$ this is
satisfied for $\lambda_0\gtrsim 7\times 10^8$~km$^{-1}$, which implies
a strong matter suppression of the collective effects, see
Fig.~\ref{fig:regions}. However, if NSI diagonal parameters are of the
order of $|\varepsilon_{\tau\tau}|\gtrsim 7.3\times
10^3/\lambda_0~({\rm km}^{-1})$ then one can avoid the matter
suppression condition.
Therefore the first NSI effect is to increase the value of
$Y_{\tau,{\rm nsi}}^{\rm eff}$ so that the $\lambda_0$ required to
have the $\mu\tau$-resonance outside $r_{\rm bip}$ is still consistent
with the presence of collective effects.
Moreover, if $\varepsilon_{\tau\tau}$ is of the order of a few
\% the condition given in Eq.~(\ref{eq:Ye_resI}) is fulfilled for the
typical values of $Y_e$ found in SNe. Thus, in the region IV we can
distinguish two subsets of parameters denoted by IVa and IVb defined
by the absence or presence of the $I$-resonance, respectively. For
defineteness we set the boundary at $\varepsilon_{\tau\tau}=10^{-2}$.

Let us first consider the IVa region. Depending on $\lambda_0$,
i.e. on the instant considered, this range of parameters implies
values of $|\varepsilon_{\tau\tau}|$ from $10^{-3}$ to $\sim 10^{-2}$.
Although values are not high enough to induce the $I$ resonance they
are sufficiently large to push the $\mu\tau$-resonance outside the
bipolar region.  The situation is therefore analogous to the one
described in Ref.~\cite{EstebanPretel:2007yq}. That means a flavor
pair transformation $\nu_e\bar\nu_e\rightarrow \nu_x\bar\nu_x$ due to
collective effects only for inverted neutrino mass hierarchy, like in
region III.  However, the final matter eigenstates depend on the
$\theta_{23}$ octant. In the middle panel of
Fig.~\ref{fig:levcros_noI} we show with solid lines the pair
conversion for $\theta_{23}$ in the first octant. In terms of matter
eigenstates, $\nu_e$ and $\bar\nu_e$ are transformed into $\nu_1^{\rm
  m}$ and $\bar\nu_1^{\rm m}$, respectively. The presence of the
$\mu\tau$-resonance in the neutrino channel leads to a difference of
$\nu_e$ with respect to region III. In the top panel of
Fig.~\ref{fig:rho_regionIII-IVa} we show with dashed lines the
evolution of $\nu_e$ as function of the distance, for
$\varepsilon_{\tau\tau}=3\times 10^{-3}$ and $\sin^2\theta_{23}=0.4$.
In the collective bipolar conversions, the excess $\epsilon$ of
$\nu_e$ over $\bar\nu_e$ remains as $\nu_2^{\rm m}$ whereas the rest
will be transformed to $\nu_1^{\rm m}$. As a consequence, the original
$\nu_e$ flux leaving the star can be written as $\rho^{\rm
  final}_{ee}=\varepsilon\sin^2\theta_{12}+\cos^2\theta_{12}$, which
in our particular case amounts to roughly 0.75.
If $\theta_{23}$ belongs to the second octant the $\mu\tau$-resonance
takes place in the antineutrino channel, see bottom panel of
Fig.~\ref{fig:levcros_noI}. The pair $\nu_e$ and $\bar\nu_e$ is driven
to the lowest-lying states, which in this case are $\nu_3^{\rm m}$ and
$\bar\nu_2^{\rm m}$, for neutrinos and antineutrinos,
respectively. Therefore, for $\nu_e$ the situation is completely
analogous to that in region III, whereas $\bar\nu_e$ leave the star as
$\bar\nu_2$.  The radial evolution of $\bar\nu_e$ for
$\varepsilon_{\tau\tau}=3\times 10^{-3}$ and $\sin^2\theta_{23}=0.6$
is displayed with dashed lines in the bottom panel of
Fig.~\ref{fig:rho_regionIII-IVa}.
It is remarkable that neither $\nu_e$ nor $\bar\nu_e$ undergo the H
resonance, and therefore are blind to the possible effect of the
outwards propagating shock
wave~\cite{Schirato:2002tg,Fogli:2003dw,Tomas:2004gr}.

It is important to notice that the same effect observed for the
different octants of $\theta_{23}$ can be obtained by fixing the
octant and changing the sign of $\varepsilon_{\tau\tau}$. This
  can be easily understood if we study the $\mu\tau$-resonance
  condition,
\begin{equation}
\label{eq:mutau_rescond}
\lambda(r)[Y^{\rm
    eff}_{\tau}+(2-Y_e)\varepsilon_{\tau\tau}] \simeq -\omega_{\rm H}
  \cos^2{\theta_{13}}\cos{2\theta_{23}}\,, 
\end{equation}
where we have neglected subleading solar terms. This condition
dictates the channel where the resonance takes place. In the standard
case this is determined only by the hierarchy of neutrino masses and
the octant of $\theta_{23}$. In the presence of NSI, though, the sign
of left-hand side of the equation depends on that of
$\varepsilon_{\tau\tau}$, what therefore affects directly the
resonance condition. As a result, the same result of
Fig.~\ref{fig:rho_regionIII-IVa} is obtained by changing the sign of
$\varepsilon_{\tau\tau}$ and the octant of $\theta_{23}$,
i.e. $\varepsilon_{\tau\tau}=-3\times 10^{-3}$ and second octant for
the panel on top and first octant in the bottom panel.
In summary, in the presence of non-universal NSI parameters with
$|\varepsilon_{\tau\tau}|\gtrsim$ $10^{-3}$, SN neutrinos are
sensitive to the octant of $\theta_{23}$, the absolute value of the
non-universal NSI parameter as well as its sign.

Finally, for higher values of the non-universal NSI parameters,
$\varepsilon_{\tau\tau}\gtrsim 10^{-2}$ (region IVb), the internal
$I$-resonance will arise. In this case one must analyze the interplay
between collective effects and the $I$-resonance. This is discussed in
the next section.

\section{Collective effects and NSI-induced $I$-resonance}
\label{sec:nsi-col}

In this section we analyze region IVb, defined by $\lambda_0\lesssim
1.4\times 10^7$~km$^{-1}$ and $\varepsilon_{\tau\tau}\gtrsim 10^{-2}$,
where both an adiabatic $I$-resonance and collective effects are
present. If the $I$-resonance is not adiabatic then neutrinos within
region IVb evolve exactly as in IVa.

As can be inferred from Fig.~\ref{fig:profiles}, one of the main
features of this scenario is that both effects happen nearly in the
same region, namely the deepest layers right above the neutrinosphere.
This means that the final result will also depend on the relative
position between the bipolar region and the location of the
$I$-resonance. Schematically two extreme scenarios can be
identified. In one case the rise  in the $Y_e$, and consequently
the $I$-resonance, takes place before the bipolar conversion region,
see $Y_e^{\rm a}$ in the bottom panel of Fig.~\ref{fig:profiles}.
In the second scenario one has first the bipolar conversion and then
neutrinos traverse the $I$-resonance, see $Y_e^{\rm b}$ in the bottom
panel of Fig.~\ref{fig:profiles}.

\subsection{First NSI $I$-resonance}
\label{subsec:nsi-col}

Let us analyze here the particular case where the $I$-resonance
happens in deeper layers than collective effects. This situation
corresponds to the $Y_e^{\rm a}$ in the bottom panel of
Fig.~\ref{fig:profiles} and, according to SN numerical simulations it
is the most likely situation.

The main consequence of an $I$-resonance in the innermost layers,
right after the neutrinosphere, is an inversion of the neutrino fluxes
entering the bipolar region. For the initial flux pattern assumed in
Sec.~\ref{sec:eoms} this implies the following new pattern after the
$I$-resonance: $F_{\nu_e} = F_{\bar\nu_e} = 0$,
$F_{\nu_\tau}=1+\epsilon$, and $F_{\bar\nu_\tau}=1$ normalized to
$F_{\bar\nu_e}$. In contrast to the standard case, under this
condition collective effects arise in the case of normal mass
hierarchy.

\begin{figure}
\begin{center}
\includegraphics[angle=0,width=0.45\textwidth]{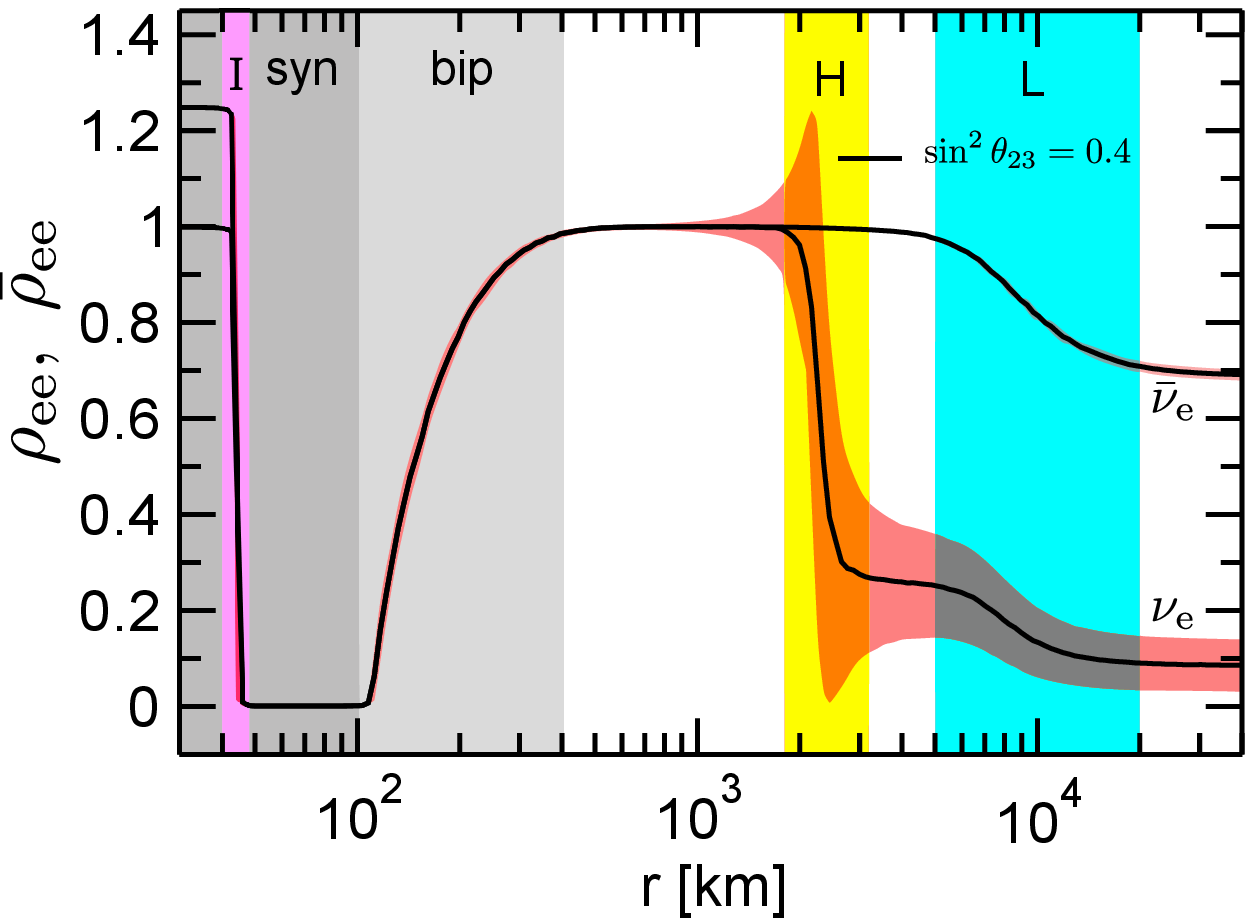}
\vskip8pt
\includegraphics[angle=0,width=0.45\textwidth]{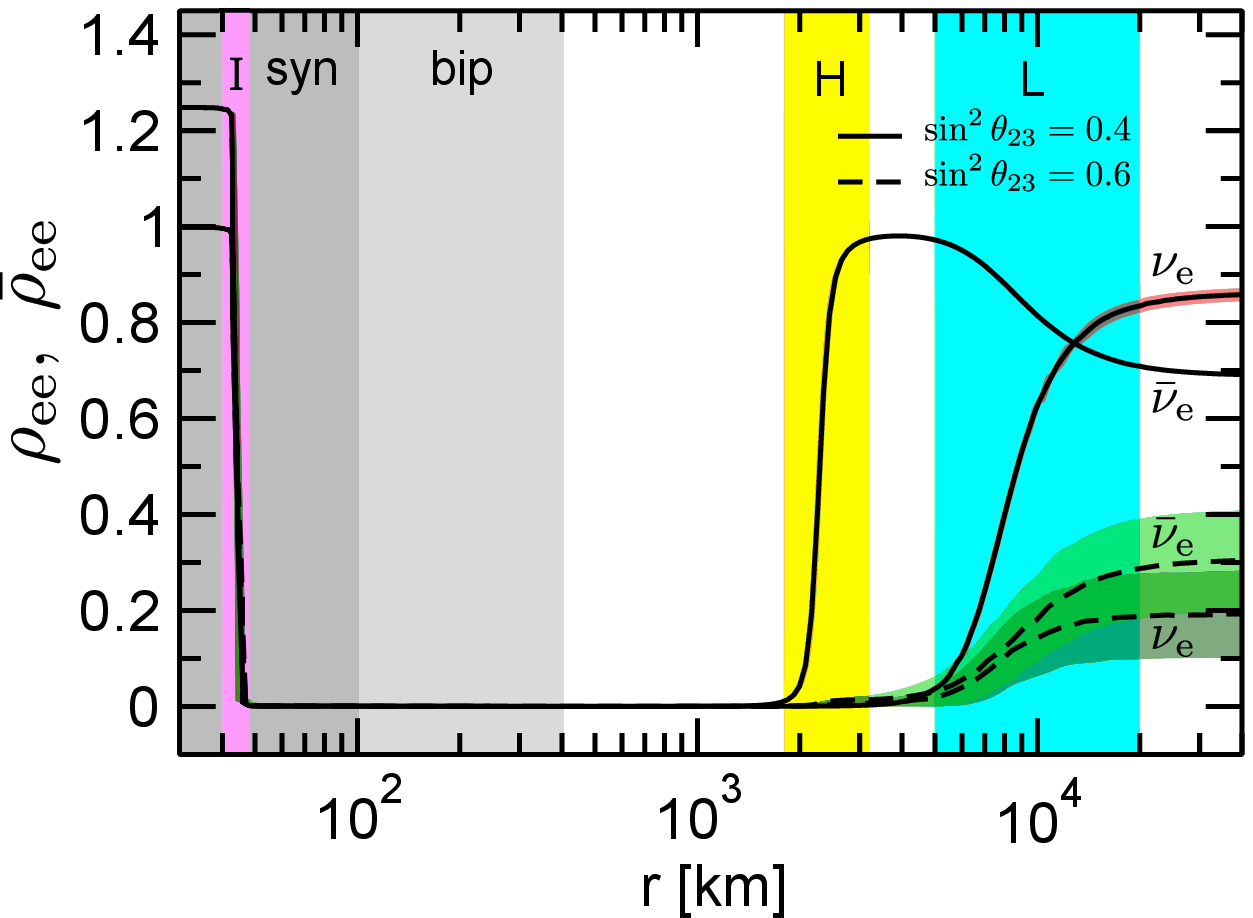}
\caption{Same as Fig.~\ref{fig:rho_regionI-II} for region IVb, for
  normal (top panel) and inverted (bottom panel) mass hierarchy. In
  the top panel it is shown the case $\sin^2\theta_{23}=0.4$, while in
  the bottom panel the case $\sin^2\theta_{23}=0.6$ is also displayed.
  In both cases $\lambda_0=4\times 10^6$~km$^{-1}$ and
  $\varepsilon_{\tau\tau}=5\times 10^{-2}$ and
  $\varepsilon_{e\tau}=10^{-3}$. The $I$-resonance is assumed to occur
  before the collective effects}
\label{fig:rho_regionIVb}
\end{center}
\end{figure}

This can be understood using the pendulum
analogy~\cite{Hannestad:2006nj} in the corresponding reduced two
flavor scenario, and keeping in mind that the bipolar conversion
drives the neutrinos to the lowest-lying states.
In the normal hierarchy the system is already created near the minimum
of the potential. Thus, in absence of the $I$-resonance collective
effects are not present. However once an adiabatic
  $I$-resonance occurs the neutrino flavor is swapped and the system
  is driven to the maximum of the potential. The original $\nu_e$
  ($\bar\nu_e$) are transformed to $\nu_\tau$ ($\bar\nu_\tau$) and, in
  terms of the vacuum level diagrams, occupy now the highest states
  right before collective effects are switched on. 
In the right panels of Fig.~\ref{fig:vaclevels} we
  show the vacuum level diagrams after an adiabatic $I-$resonance has
  occurred. In the case of normal mass hierarchy (bottom right panel)
  bipolar effects act taking $\nu_e$ ($\bar\nu_e$) to the lowest-lying
  states $\nu_1'$ ($\bar\nu_1'$), which in this case
  do not depend on the $\theta_{23}$ octant.  This bipolar conversion
  can be also seen in term of matter eigenstates as dashed arrows in
  the top panel of Fig.~\ref{fig:levcros_I}.  As a result both the
  $I$-resonance and the induced collective effects basically cancel
  each other. In the top panel of Fig.~\ref{fig:rho_regionIVb} we show
  the radial evolution of $\nu_e$ and $\bar\nu_e$. This cancellation
  between the $I$-resonance and collective effects is complete for
  $\bar\nu_e$, which leave as $\bar\nu_1$, but not for $\nu_e$: its
  excess $\epsilon$ over $\bar\nu_e$ remains as $\nu_3$ ($\nu_2$) for
  $\theta_{23}$ in the first (second) octant. Therefore for a
  monochromatic neutrino flux we obtain, after propagating through the
  outer resonances, $\rho^{\rm
    final}_{ee}=\epsilon\sin^2\theta_{12}+\sin^2\theta_{13}\approx
  0.08$ for both $\theta_{23}$ octants, see top panel of
  Fig.~\ref{fig:rho_regionIVb}, instead of simply
  $\sin^2\theta_{13}=10^{-2}$ as in top panel of
  Fig.~\ref{fig:rho_regionI-II}.  Hence, by comparing the top panels
  of Figs.~\ref{fig:rho_regionI-II} and~\ref{fig:rho_regionIVb} one
  realizes that, except for the excess $\epsilon$, the situation for
  normal mass hierarchy is basically the same as in regions I, III,
  and IVa.

\begin{figure}
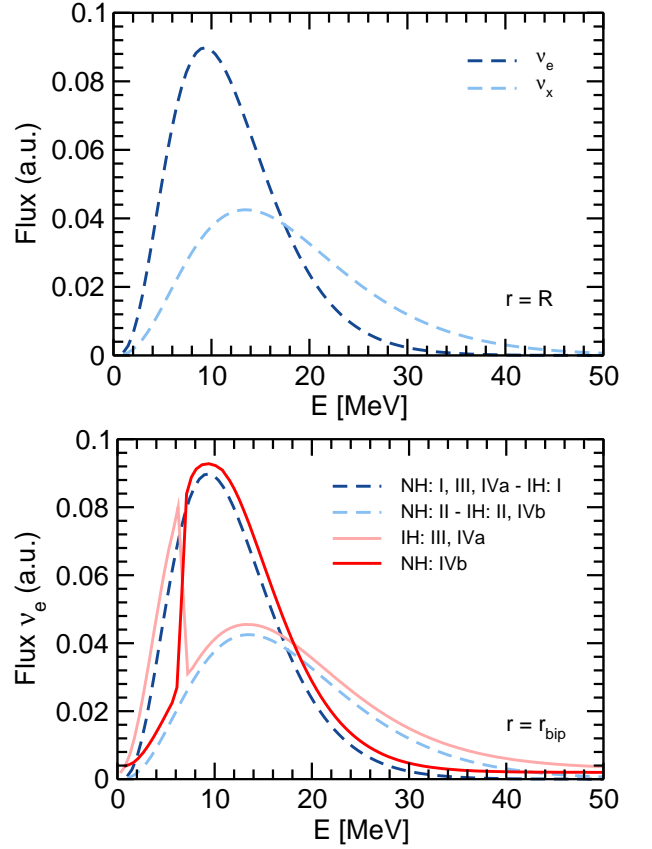

\begin{center}
\includegraphics[angle=0,width=0.45\textwidth]{fig9a.eps}
\includegraphics[angle=0,width=0.45\textwidth]{fig9b.eps}
\caption{Top: $\nu_e$ and $\nu_x$  fluxes as emitted at the
  neutrinosphere. Bottom: $\nu_e$ flux after the bipolar region for
  different cases. In the region IVb it is assumed that the $I$-resonance 
  occurs before than the bipolar conversion. Note a 
  depletion in the flux of low-energy neutrinos, instead of usual spectral split.
}
\label{fig:spectralsplit}
\end{center}
\end{figure}

While this is true in the monoenergetic case
a specific signature can be observed if we do not restrict
ourselves to that case but consider the whole energy
spectrum. The top panel of Fig.~\ref{fig:spectralsplit} displays the 
$\nu_e$ and $\nu_x$ fluxes at the neutrinosphere, $f^{\rm R}_{\nu_e}$
and $f^{\rm R}_{\nu_x}$. We have assumed the parameterization given in
Ref.~\cite{Keil:2002in}, 
\begin{eqnarray}
  f^R_{\nu_\alpha}(E) = C_{\nu_\alpha}
  \left(\frac{E}{\langle{E_{\nu_\alpha}}\rangle}\right)^{\beta_{\nu_\alpha}-1} 
  \exp\left(-\beta_{\nu_\alpha}\frac{E}{\langle{E_{\nu_\alpha}}\rangle}\right) \,,
\label{eq:flux-Gal}
\end{eqnarray}
with $\langle E_{\nu_e}\rangle = 12$~MeV, $\langle
E_{\nu_e}\rangle=15$~MeV, $\langle E_{\nu_x}\rangle=18$~MeV,
$\beta_{\nu_e}=5,~\beta_{\bar\nu_e}=4.5$ and $\beta_{\nu_x}=4$. The
normalization $C_{\nu_\alpha}$ has been chosen such that $F^{\rm
  R}_{\bar\nu_e}\equiv \int f^{\rm R}_{\bar\nu_e}(E){\rm d}E=1$,  $F^{\rm
  R}_{\nu_e} = 1+\kappa\epsilon$ and $F^{\rm  R}_{\nu_x} = 1-\kappa$,
with $\kappa=0.15$.

In the bottom panel of Fig.~\ref{fig:spectralsplit} we show the
$\nu_e$ fluxes after the bipolar region. In solid dark red lines we
represent the case of normal mass hierarchy in region IVb. By
comparing the two panels one sees how the conversion
$\nu_e\rightarrow\nu_x$ takes place only at low energies. This is
exactly the opposite of what happens for the standard case (inverted
mass hierarchy in region III), shown as solid light red lines, where
the untransformed flux concentrates at low
energies~\cite{Raffelt:2007cb,Raffelt:2007xt}.
For completeness we show also the other cases. In the region of
parameters I, III and IVa there is neither collective effects nor
$I$-resonance for normal hierarchy, then the fluxes after the
bipolar region coincide with the initial ones, $f_{\nu_e}=f_{\nu_e}^R$
and $f_{\nu_x}=f_{\nu_x}^R$. In region II, the $I$-resonance implies a
complete conversion $\nu_e\rightarrow \nu_x$, which leads to a spectral
swap, $f_{\nu_e}=f_{\nu_x}^R$ and $f_{\nu_x}=f_{\nu_e}^R$. 

The case of inverted mass hierarchy is more subtle. According to the
previous discussion one would expect no collective effects after
neutrinos traverse the $I$-resonance.  The system starts its evolution
near the maximum of the potential and, in the absence of NSI, the bipolar
conversions would take it to the minimum. What the $I$-resonance is
doing in this language by swapping the flavor eigenstates is to take
the system to the minimum of the potential before any collective
effects can arise. The new stable situation prevents bipolar
conversions, leaving the system unchanged until the outer resonances
are reached. 
And this is indeed what happens if $\theta_{23}$ lies in the first
octant. In terms of the vacuum levels diagram the
  original $\nu_e$ ($\bar\nu_e$) are transformed after the $I$-resonance into
  $\nu_\tau$ ($\bar\nu_\tau$), which are already the lowest-lying
  states, see the top right panel in Fig.~\ref{fig:vaclevels}.  Hence
  no collective effects take place, and, as can be seen in the middle
  panel of Fig.~\ref{fig:levcros_I} after traversing the outer
  resonances $\nu_e$ and $\bar\nu_e$ leave the star as $\nu_1$ and
  $\bar\nu_1$. The radial evolution is shown with solid lines in the
  bottom panel of Fig.~\ref{fig:rho_regionIVb}.  

However if $\theta_{23}$ lies in the second octant things are
different. As in the previous case, the $I$-resonance converts
neutrinos created initially as $\nu_e$ and $\bar\nu_e$ into $\nu_\tau$
and $\bar\nu_\tau$, respectively. But these states are
  now the highest states, $\nu_2'$ and $\bar\nu_2'$.  This means, that
  in contrast to the first-octant case, when the neutrinos traverse
  the bipolar regime they will be driven to the lowest-lying states,
  i.e. $\nu_2'\bar\nu_2'\rightarrow\nu_3'\bar\nu_3'$, see top right panel
  of Fig.~\ref{fig:vaclevels}.  In the bottom panel of
  Fig.~\ref{fig:rho_regionIVb} we show with dashed lines the evolution
  of $\nu_e$ and $\bar\nu_e$ as function of distance for
  $\sin^2\theta_{23}=0.6$. The bipolar conversion can not be directly
  seen as it occurs between states containing explicitly neither
  $\nu_e$ nor $\bar\nu_e$.
At the end, as shown in the bottom panel of Fig.~\ref{fig:levcros_I},
the original $\bar\nu_e$ leave the SN as $\bar\nu_2$, see bottom panel
of Fig.~\ref{fig:rho_regionIVb}. In the case of $\nu_e$ the excess
$\epsilon$ over $\bar\nu_e$ remains as $\nu_2$ whereas the
rest is transformed to $\nu_3$. Therefore for a monoenergetic
flux we find $\rho^{\rm final}_{ee} = \epsilon
\cos^2\theta_{12}+\sin^2\theta_{13}\approx 0.2$. By comparing the
bottom panel of Figs.~\ref{fig:rho_regionIVb}
and~\ref{fig:rho_regionIII-IVa} one realizes that this case is
analogous to IVa.
Since the collective effects do not affect $\nu_e$ and $\bar\nu_e$
directly, considering neutrinos with an energy spectrum, one expects
simply a complete swap of spectra, $f_{\nu_e}=f_{\nu_x}^R$ and
$f_{\nu_x}=f_{\nu_e}^R$ after the bipolar region,
like in scenario II, see bottom panel of Fig.~\ref{fig:spectralsplit}.

The final conclusion is that the propagation of neutrinos with
parameters in region IV is practically independent of whether the
$I$-resonance is present (IVb) or not (IVa). The main consequence of
the $I$-resonance is to swap the $\nu_e$ spectrum at low energies
and only in the case of normal mass hierarchy, in contrast to the
transformation of high-energy $\nu_e$ happening for inverted mass
hierarchy in the absence of $I$-resonance.

\subsection{First Collective}
\label{subsec:col-nsi}

For completeness we have also considered the possibility that the
bipolar conversion takes place before neutrinos traverse the
$I$-resonance.  This situation corresponds schematically to the
$Y_e^{\rm b}$ profile in the bottom panel of Fig.~\ref{fig:profiles}.

\begin{figure}
\begin{center}
\includegraphics[angle=0,width=0.45\textwidth]{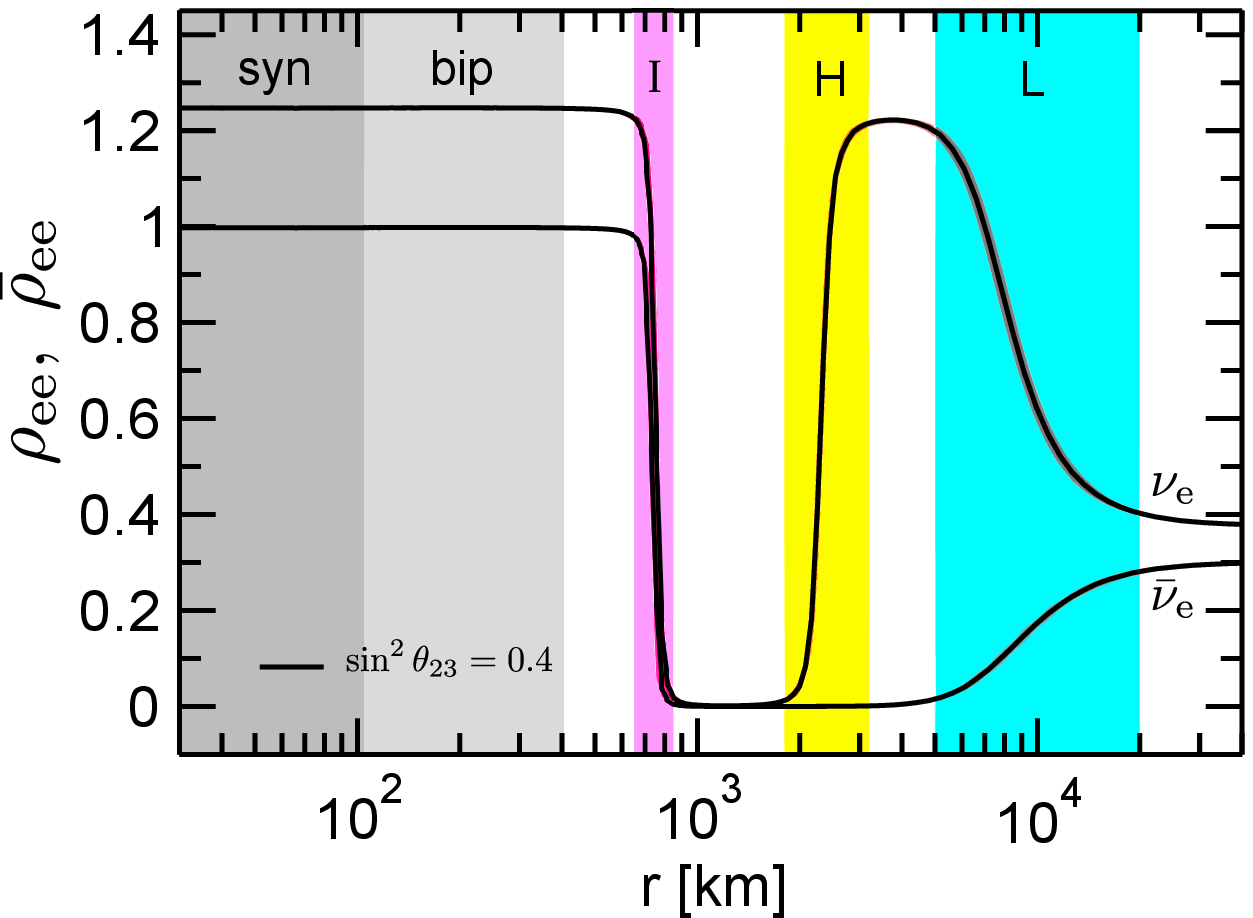}
\includegraphics[angle=0,width=0.45\textwidth]{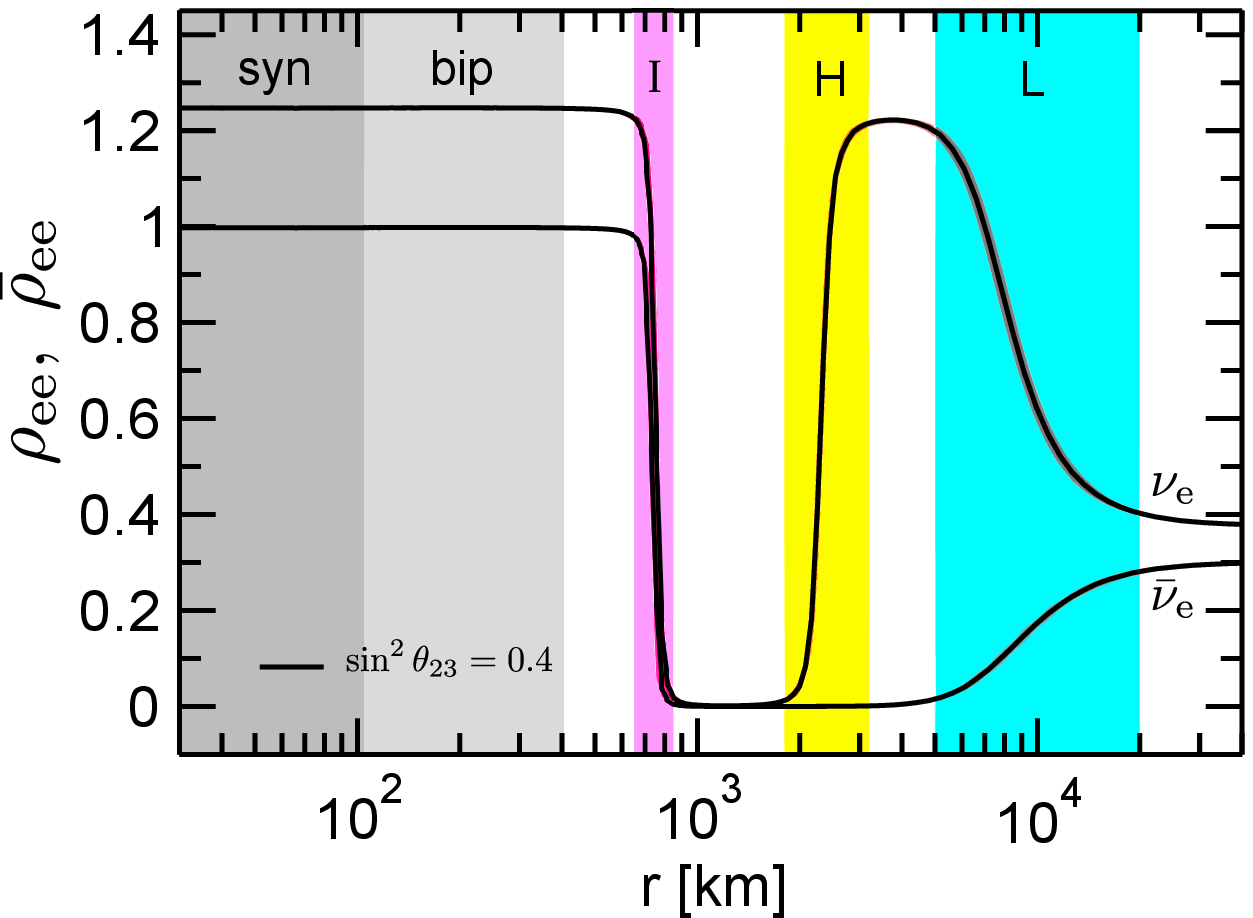}
\caption{Same as Fig.~\ref{fig:rho_regionIVb} but assuming that the
  collective effects take place before the $I$-resonance.}
\label{fig:rho_regionIVbbis}
\end{center}
\end{figure}

The case of normal mass hierarchy is completely analogous to the one
of region II, that is, absence of collective effects and
$I$-resonance. The $\nu_e$ and $\bar\nu_e$ are created as $\nu_2^{\rm
  m}$ and $\bar\nu_2^{\rm m}$, respectively. Therefore, if all
resonances involved, $I,~\mu\tau,~H,$ and $L$, are adiabatic then they
leave the SN as $\nu_2$ and $\bar\nu_2$, respectively, see top panel
of Fig.~\ref{fig:rho_regionIVbbis}. The result is identical to the one
shown with dashed lines in the top panel of
Fig.~\ref{fig:rho_regionI-II}.

The situation with inverted mass hierarchy depends significantly on
the $\theta_{23}$ octant. Rotating the matter term away $\nu_e$ and
$\bar\nu_e$ are created as the intermediate states $\nu_1'$ and
$\bar\nu_1'$. Collective effects drive them to the lowest-lying states
$\nu_3'$ and $\bar\nu_3'$. However the corresponding matter eigenstates
are different depending on whether $\theta_{23}$ belongs to the
first or to the second octant. In the first case, most of $\nu_e$ and
$\bar\nu_e$ end up as $\nu_2^{\rm m}$ and $\bar\nu_3^{\rm m}$ before
crossing the $I$-resonance, see solid arrows in the middle panel of
Fig.~\ref{fig:levcros_I}. The excess $\epsilon$ of $\nu_e$ stays as
$\nu_1^{\rm m}$. As a consequence, the final $\nu_e$ and $\bar\nu_e$
fluxes, normalized to the initial $\bar\nu_e$ one, are $\rho^{\rm
  final}_{ee}=\epsilon\cos^2\theta_{12}+\sin^2\theta_{12}$ and
$\bar\rho^{\rm final}_{ee}=\sin^2\theta_{13}$, respectively. See
solid lines in the bottom panel of
Fig.~\ref{fig:rho_regionIVbbis}. Except for the excess $\epsilon$ in
$\nu_e$ the net result is a cancellation of the collective effects and
the $I$-resonance, leading to a similar result as in region I (solid
lines in bottom panel of Fig.~\ref{fig:rho_regionI-II}).
Qualitatively, the main difference shows up in the $\nu_e$
spectrum. The initial collective effects induces a ``standard'' swap
only at high energies. Nevertheless, as neutrinos cross the
$I$-resonance this split turns into an inverse one, with a swap at low
energies. The final result right after the $I$-resonance is analogous
to the case of normal mass hierarchy and the $I$-resonance happening
first, discussed in Sec.~\ref{subsec:nsi-col} and displayed with dark
red solid lines in the bottom panel of Fig.~\ref{fig:spectralsplit}.

If $\theta_{23}$ lies in the second octant then most of $\nu_e$ and
$\bar\nu_e$ end up as $\nu_3^{\rm m}$ and $\bar\nu_2^{\rm m}$. As can
be seen in the solid lines in the bottom panel of
Fig.~\ref{fig:levcros_I} these neutrinos will not traverse the
$I$-resonance, except the excess $\epsilon$ of $\nu_e$, which stays as
$\nu_1^{\rm m}$. These neutrinos will be basically blind to the $I$-,
$H$-, and $L$-resonances. The final fluxes will be therefore
$\rho^{\rm final}_{ee}=\sin^2\theta_{13}+\epsilon\cos^2\theta_{12}$
and $\bar\rho^{\rm final}_{ee}=\sin^2\theta_{12}$.
This case is represented with dashed lines in the bottom panel of
Fig.~\ref{fig:rho_regionIVbbis}. 
In the end the final evolution turns out to be similar to that in
region IVb.

\section{Discussion}
\label{sec:discussion}

In the previous sections we have studied the consequences of NSI on
the neutrino propagation through the SN envelope 
taking into account the presence of a neutrino background.
We have analyzed the different situations in terms of the
non-universal NSI parameter $\varepsilon_{\tau\tau}$ and the density
at the neutrinosphere $\lambda_0$. Depending on their values we were
able to identify four extreme regions of the parameters where the
evolution of the neutrinos have a specific pattern.

\begin{figure*}
\begin{center}
\includegraphics[angle=0,width=0.8\textwidth]{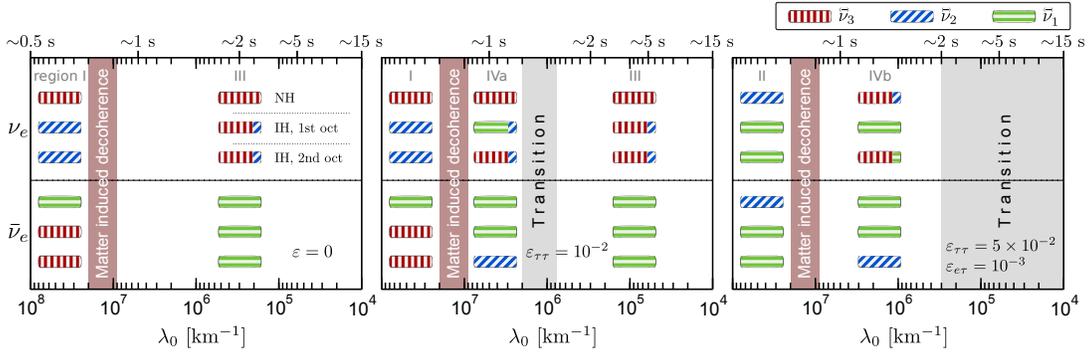}
\caption{
Relationship between (anti)neutrinos created as $\nu_e$
(upper half part) and $\bar\nu_e$ (lower half part) and the matter
eigenstates at the SN surface as function of time for different
neutrino mass and mixing schemes and for given values of
$\varepsilon_{\tau\tau}=0$ (left), $10^{-2}$ (middle), and
$5\times 10^{-2}$ (right).
}
\label{fig:finalscheme}
\end{center}
\end{figure*}

In a realistic situation, though, we expect to find a combination of
these situations depending on the instant considered. As mentioned in
Sec.~\ref{sec:eoms} one expects the value of $\lambda_0$ to decrease
with time as the explosion goes on.
Therefore it is important to look for a time dependence in the
neutrino propagation for given values of the NSI parameters.  In
Fig.~\ref{fig:finalscheme} we show  the
  composition in terms of mass eigenstates of neutrinos created as
  $\nu_e$ (upper half of the panels) and $\bar\nu_e$ (lower half of
  the panels) at the neutrinosphere, when they leave the SN, as
function of time for different neutrino mass and mixing schemes and
for a given value of $\varepsilon_{\tau\tau}$.  The evolution in time
shown in each panel is equivalent to consider Fig.~\ref{fig:regions},
fix a value in the $x$ axis corresponding to some
$\varepsilon_{\tau\tau}$ and following vertically towards lower values
of $\lambda_0$.  Depending on $\varepsilon_{\tau\tau}$ and the instant
considered one can distinguish different regions separated by vertical
bands denoting transition phases. The position and the size of these
transition bands are not constant but depend on time, as $\mu_0$ and
$\epsilon$ do. Nevertheless, unless multiangle decoherence is
triggered (by e.g. a strong reduction of
$\epsilon$~\cite{EstebanPretel:2007ec}), the sequence of different
regimes undergone by the neutrinos is not expected to change
drastically.

Let us first discuss the antineutrino case. In the
  left panel we consider the standard case,
  i.e. $\varepsilon_{\tau\tau}=0$. On the left we have early times
(or large $\lambda_0$), which corresponds to the region I in
Fig.~\ref{fig:regions}. From the previous discussion,
  we know that for such a case and normal mass hierarchy (upper box)
  an antineutrino created as $\bar\nu_e$ leaves the star as
  $\bar\nu_1$ whereas in inverted mass hierarchy they escape as
  $\bar\nu_3$ due to the adiabatic $H$-resonance for both first and second
  $\theta_{23}$ octants (middle and lower boxes, respectively), see
  Figs.~\ref{fig:levcros_noI} and~\ref{fig:rho_regionI-II}.
At later times, $\lambda_0$ becomes smaller and matter can not
suppress collective effects any longer, i.e. neutrinos enter region
III. These affect only in the inverted mass hierarchy case
``canceling'' the $H$-resonance conversion and making
  the initial $\bar\nu_e$ escape as $\bar\nu_1$. There is then a time
dependence in the survival $\bar\nu_e$ probability for inverted mass
hierarchy but not for the normal one. As can be seen in the panel this
behavior does not depend on the $\theta_{23}$ octant, see
Figs.~\ref{fig:levcros_noI} and~\ref{fig:rho_regionIII-IVa}. In terms
of $\bar\nu_e$ survival probabilities there is then a transition from
$\sin^2\theta_{13}\approx 0$ at early times to
$\cos^2\theta_{12}\approx 0.7$ at later times, the details depending
on the specific time evolution of $\lambda(r)$.

Let us now take the middle panel with
$\varepsilon_{\tau\tau}=3\times 10^{-3}$. The situation at early times
is the same as in the previous panel, described by the region
I. However at intermediate times the situation changes in the case of
inverted mass hierarchy. Now the NSI parameters make the evolution go
through region IVa before entering eventually region III. The
$\mu\tau$-resonance is pushed outside the bipolar region and then the
degeneracy between the two $\theta_{23}$ octants is broken: for
$\theta_{23}$ in the first octant $\bar\nu_e$ leaves as
$\bar\nu_1$ whereas for the second octant they escape as
$\bar\nu_2$, see Figs.~\ref{fig:levcros_noI}
and~\ref{fig:rho_regionIII-IVa}. At later times $\lambda_0$ further
decreases and the $\mu\tau$-resonance contracts to deeper layers
within $r_{\rm syn}$. Hence neutrinos cross to region III and the
$\theta_{23}$ octant degeneracy is restored.  Concerning the
$\bar\nu_e$ survival probability, as before, there is a transition
  from $\sin^2\theta_{13}\approx 0$ to $\cos^2\theta_{12}\approx
  0.7$. For $\theta_{23}$ in the first octant it is direct, whereas if
  $\theta_{23}$ lies in the second octant it goes from
  $\sin^2\theta_{13}\approx 0$ through $\sin^2\theta_{12}\approx
  0.3$.  As it was discussed in Sec.~\ref{subsec:regionIV} an
analogous effect would arise for a fixed $\theta_{23}$ and different
signs of $\varepsilon_{\tau\tau}$. For NH there is no time
  dependence.

  Finally, we consider the case where the NSI parameters are large
  enough, $\varepsilon_{\tau\tau}=5\times 10^{-2}$, to induce the
  $I$-resonance, right panel of
    Fig.~\ref{fig:finalscheme}. Now, at early times neutrino
  propagation follows the prescription given in region II. For normal
  mass hierarchy the original $\bar\nu_e$ leave the
    star as $\bar\nu_2$, whereas for inverted they escape as
    $\bar\nu_1$ for both octants, see Figs.~\ref{fig:levcros_I}
  and~\ref{fig:rho_regionIVb}. After this phase neutrinos enter the
  region IVb and collective effects {\em are switched
    on}. For NH they cancel the $I$-induced resonant conversion,
  whereas for IH they break the degeneracy of the $\theta_{23}$ octant
  for inverted mass hierarchy.

  The bottom line is that if $|\varepsilon_{\tau\tau}|\gtrsim$ a
  $10^{-3}$ neutrinos cross the region IV during some seconds, and
  this could help disentangle the $\theta_{23}$ octant. If the octant
  were known one could obtain information about the sign of the
  non-universal NSI parameters.

  In the upper half part of the panels we show the
    same kind of plots but for neutrinos. The main difference with
    respect to antineutrinos is that in the presence of collective
    effects $\nu_e$ are not fully converted like $\bar\nu_e$. Some
    fraction of them, corresponding to the excess over $\bar\nu_e$,
    remains unaffected. This excess is represented in
    Fig.~\ref{fig:finalscheme} as a small portion at the
    right hand side of the corresponding boxes.  As discussed, this
    excess of $\nu_e$ translates to a spectral split, so that the
    flavor spectral swap happens only for some energies. Whether these
    correspond to the low-energy tail or high-energy tail of the
    initial spectrum depends on the neutrino oscillation parameters,
    see bottom panel of Fig.~\ref{fig:spectralsplit}.  Therefore one
    could hope to use this additional information to break possible
    degeneracies between different mass and mixing schemes and
    different values of the NSI parameters.
It is important to note that the results here presented are based on
the assumption of a particular choice of the initial neutrino fluxes,
with the hierarchy $F^R_{\nu_e} > F^R_{\bar\nu_e} >
F^R_{\nu_x}$. However, as shown in Ref.~\cite{Dasgupta:2009mg}, a
different pattern for the initial fluxes could lead to a more
complicated structre of the final spectrum with multilpe spectral
splits.

Throughout the analysis we have concentrated on the case of NSI with
$d$ quarks. Nevertheless most of the results here presented can be
generalized to the case of $u$ quarks and $e$. In the first case the
only effect is to shift the position of the $I$-resonance, since the
resonance condition is modified to
$Y_e^I=\varepsilon_{\tau\tau}/(1-\varepsilon_{\tau\tau})$
~\cite{EstebanPretel:2007yu}. For the case of NSI involving electrons
the $I$-resonance is absent. But nevertheless its contribution to
increase the value of $Y^{\rm eff}_{\rm \tau,nsi}$ would also make the
neutrino propagation highly sensitive to the $\theta_{23}$ octant and
to its own sign, exactly as in the case of $d$ quark.

Last but not least we briefly comment on the possibility to observe
the different regimes analyzed. This possibility will be hampered by
several uncertainties inherent in SN neutrinos. One is the lack of
knowledge on the exact matter profile traversed by the outgoing
neutrinos. In our study we assumed a simple power law given in
Eq.~(\ref{eq:lambda(r)}). This density profile will be significantly
distorted by the passage of the shock wave responsible for ejecting
the whole SN
envelope~\cite{Schirato:2002tg,Fogli:2006xy,Friedland:2006ta}. One of
the main effect will be to destroy the adiabaticity of the
$H$-resonance, which was assumed in our study. This effect, though, is
not always present but depends on the neutrino parameters.  Therefore,
far from being a problem, the time and energy dependence modulation
introduced in the spectra could further help disentangle between the
different scenarios here
considered~\cite{Fogli:2003dw,Tomas:2004gr,EstebanPretel:2007yu,Gava:2009pj}.

Another important source of uncertainties is our ignorance of the
exact initial fluxes $f^{\rm R}_{\nu_\alpha}(E)$. Although the initial
fluxes during the first stage of the explosion, the neutronization
burst, are rather model independent~\cite{Kachelriess:2004ds}, the
expected number of events is very low. Most of the signal is generated
later, during the accretion and cooling phases. The spectral features
observed in the numerical simulations depend strongly on the
properties of the SN.
It is therefore necessary to set up strategies combining different
observables to be able to pin down the underlying neutrino properties
independently of the initial fluxes. For example spectral modulations
are expected if neutrinos cross the Earth before being
detected~\cite{Dasgupta:2008my,Lunardini:2001pb,Dighe:2003be,Dighe:2003jg,Dighe:2003vm}. One
may also take advantage of the time dependence of the matter profiles
$\lambda(r)$ and $Y_e(r)$ themselves~\cite{EstebanPretel:2007yu}.

\section{Summary}
\label{sec:summary}

Here we have considered the effect of non-standard neutrino
interactions on the propagation of neutrinos through the SN
envelope. We adopted a realistic three-neutrino framework, properly
taking into account the presence of a neutrino background.
We have found that for non-universal NSI parameters exceeding
$10^{-3}$, i.e.  $|\varepsilon_{\tau\tau}|\gtrsim$ $10^{-3}$, the
neutrino propagation becomes for some time sensitive to the
$\theta_{23}$ octant and the sign of $\varepsilon_{\tau\tau}$.
In particular, for $\varepsilon_{\tau\tau}\gtrsim 10^{-2}$ an internal
$I$-resonance may arise independently of the matter density. For
typical values found in simulations this takes place in the same
dense-neutrino region above the neutrinosphere where collective
effects occur, in particular during the synchronization regime. This
resonance may lead to an exchange of the neutrino fluxes entering the
bipolar regime. The main implications are (i) bipolar conversion
taking place for normal neutrino mass hierarchy and (ii) a
transformation of the flux of low-energy $\nu_e$, instead of usual
spectral swap.

\section*{Acknowledgments}

The authors wish to thank S. Pastor and M. A. T\'ortola for fruitful
discussions.  This work is supported by the Spanish grants
FPA2008-00319/FPA and PROMETEO/2009/091 and by the DFG (Germany) under
grant SFB-676. A.E. is supported by a FPU grant from the Spanish
Government.

\section*{References}

\def\baselinestretch{1.2}
\bibliographystyle{h-physrev4} 

\begin{thebibliography}{10}

\bibitem{Eguchi:2002dm}
KamLAND collaboration, K.~Eguchi {\em et~al.},
\newblock Phys. Rev. Lett. {\bf 90}, 021802 (2003), [hep-ex/0212021];
T.~Araki {\em et~al.},
\newblock Phys. Rev. Lett. {\bf 94}, 081801 (2005).

\bibitem{Ahn:2006zz}
K2K collaboration, M.~H. Ahn,
\newblock hep-ex/0606032.

\bibitem{Adamson:2008zt}
MINOS collaboration, P.~Adamson {\em et~al.},
\newblock 0806.2237.

\bibitem{:2008zn}
 Super-Kamiokande Collaboration, J.~P.~Cravens {\it et al.} 
  Phys.\ Rev.\  D {\bf 78}, 032002 (2008)
  [arXiv:0803.4312 [hep-ex]].
 S.~Fukuda {\it et al.},
  Phys.\ Lett.\  B {\bf 539}, 179 (2002)
  [arXiv:hep-ex/0205075].

\bibitem{Aharmim:2008kc}
SNO collaboration, B.~Aharmim {\em et~al.},
\newblock Phys. Rev. Lett. {\bf 101}, 111301 (2008), [arXiv:0806.0989];
\newblock Phys. Rev. {\bf C72}, 055502 (2005), [nucl-ex/0502021];
S.~N. Ahmed {\em et~al.},
\newblock Phys. Rev. Lett. {\bf 92}, 181301 (2004), [nucl-ex/0309004].

\bibitem{abdurashitov:2002nt}
GNO collaboration,  M.~Altmann {\it et al.},
  Phys.\ Lett.\  B {\bf 490}, 16 (2000)
  [arXiv:hep-ex/0006034],
SAGE collaboration, J.~N. Abdurashitov {\em et~al.},
\newblock J. Exp. Theor. Phys. {\bf 95}, 181 (2002), [astro-ph/0204245].

\bibitem{Collaboration:2008mr}
Borexino Collaboration,
\newblock arXiv:0808.2868.

\bibitem{Ashie:2005ik}
Super-Kamiokande collaboration, Y.~Ashie {\em et~al.},
\newblock Phys. Rev. {\bf D71}, 112005 (2005), [hep-ex/0501064].

\bibitem{Ashie:2004mr}
Super-Kamiokande collaboration, Y.~Ashie, {\em et~al.},
\newblock hep-ex/0404034.

\bibitem{Pakvasa:2003zv}
S.~Pakvasa and J.~W.~F. Valle,
\newblock hep-ph/0301061,
\newblock Proc. of the Indian National Academy of Sciences on Neutrinos, Vol.
  70A, No.1, p.189 - 222 (2004), Eds. D. Indumathi, M.V.N. Murthy and G.
  Rajasekaran.

\bibitem{Schwetz:2008er}
T.~Schwetz, M.~Tortola and J.~W.~F. Valle,
\newblock New J. Phys. {\bf 10}, 113011 (2008), [arXiv:0808.2016];
for a recent review with references to all groups see,
M.~Maltoni et al, 
\newblock New J. Phys. {\bf 6}, 122 (2004) [hep-ph/0405172v6] .

\bibitem{schechter:1980gr}
J.~Schechter and J.~W.~F. Valle,
\newblock Phys. Rev. {\bf D22}, 2227 (1980).

\bibitem{Valle:2006vb}
J.~W.~F. Valle,
\newblock J. Phys. Conf. Ser. {\bf 53}, 473 (2006), [hep-ph/0608101],
\newblock Review based on lectures at the Corfu Summer Institute on Elementary
  Particle Physics in September 2005.

\bibitem{Schechter:1981hw}
J.~Schechter and J.~W.~F. Valle,
\newblock Phys. Rev. {\bf D24}, 1883 (1981),
\newblock Err. D25, 283 (1982); for recent analyses see
  O.~G.~Miranda et al,
  Nucl.\ Phys.\  B {\bf 595}, 360 (2001)
  [arXiv:hep-ph/0005259];
  Phys.\ Rev.\  D {\bf 70}, 113002 (2004)
  [arXiv:hep-ph/0406066];
  Phys.\ Rev.\ Lett.\  {\bf 93}, 051304 (2004)
  [arXiv:hep-ph/0311014].

\bibitem{Wolfenstein:1977ue}
L.~Wolfenstein,
\newblock Phys. Rev. {\bf D17}, 2369 (1978).

\bibitem{MS}
{Mikheev, S. P. and Smirnov, A. Yu.},
\newblock (Editions Fronti\`eres, Gif-sur-Yvette, 1986, p.355.),
\newblock 86 Massive Neutrinos in Astrophysics and Particle Physics,
  Proceedings of the Sixth Moriond Workshop, ed. by {Fackler}, O. and {Tran
  Thanh Van}, J.

\bibitem{Valle:1987gv}
J.~W.~F. Valle,
\newblock Phys. Lett. {\bf B199}, 432 (1987).

\bibitem{mohapatra:1986bd}
R.~N. Mohapatra and J.~W.~F. Valle,
\newblock Phys. Rev. {\bf D34}, 1642 (1986).

\bibitem{Bernabeu:1987gr}
J.~Bernabeu {\em et~al.},
\newblock Phys. Lett. {\bf B187}, 303 (1987).

\bibitem{Branco:1989bn}
G.~C. Branco, M.~N. Rebelo and J.~W.~F. Valle,
\newblock Phys. Lett. {\bf B225}, 385 (1989);
N.~Rius and J.~W.~F. Valle,
\newblock Phys. Lett. {\bf B246}, 249 (1990).

\bibitem{Deppisch:2004fa}
F.~Deppisch and J.~W.~F. Valle,
\newblock Phys. Rev. {\bf D72}, 036001 (2005), [hep-ph/0406040].

\bibitem{Malinsky:2005bi}
M.~Malinsky, J.~C. Romao and J.~W.~F. Valle,
\newblock Phys. Rev. Lett. {\bf 95}, 161801 (2005), [hep-ph/0506296].

\bibitem{Hirsch:2009mx}
  M.~Hirsch, S.~Morisi and J.~W.~F.~Valle,
  Phys.\ Lett.\  B {\bf 679}, 454 (2009)
  [arXiv:0905.3056 [hep-ph]].

\bibitem{Ibanez:2009du}
  D.~Ibanez, S.~Morisi and J.~W.~F.~Valle,
  Phys.\ Rev.\  D {\bf 80}, 053015 (2009)
  [arXiv:0907.3109 [hep-ph]].

\bibitem{He:2009xd}
X.-G. He and E.~Ma,
\newblock  arXiv:0907.2737.

\bibitem{Zee:1980ai}
A.~Zee,
\newblock Phys. Lett. {\bf B93}, 389 (1980).

\bibitem{Babu:1988ki}
K.~S. Babu,
\newblock Phys. Lett. {\bf B203}, 132 (1988).

\bibitem{AristizabalSierra:2007nf} For recent work on low-scale
  generation of neutrino masses and/or non-standard interactions see,
  for example, D.~Aristizabal Sierra, M.~Hirsch and S.~G.~Kovalenko,
  Phys.\ Rev.\  D {\bf 77}, 055011 (2008)
  [arXiv:0710.5699 [hep-ph]];
 T.~Ohlsson, T.~Schwetz and H.~Zhang,
  arXiv:0909.0455 [hep-ph];
A.~Abada, C.~Biggio, F.~Bonnet, M.~B.~Gavela and T.~Hambye,
  JHEP {\bf 0712}, 061 (2007)
  [arXiv:0707.4058 [hep-ph]];
M.~Malinsky, T.~Ohlsson and H.~Zhang,
  Phys.\ Rev.\  D {\bf 79}, 011301 (2009)
  [arXiv:0811.3346 [hep-ph]];
J.~Chakrabortty, A.~Dighe, S.~Goswami and S.~Ray,
  Nucl.\ Phys.\  B {\bf 820}, 116 (2009)
  [arXiv:0812.2776 [hep-ph]];
  Y.~Liao, J.~Y.~Liu and G.~Z.~Ning,
  Phys.\ Rev.\  D {\bf 79}, 073003 (2009)
  [arXiv:0902.1434 [hep-ph]];
Z.~z.~Xing and S.~Zhou,
  arXiv:0906.1757 [hep-ph];
 F.~Bonnet, D.~Hernandez, T.~Ota and W.~Winter,
  arXiv:0907.3143 [hep-ph].

\bibitem{Bazzocchi:2009kc}
F.~Bazzocchi et al,
\newblock  arXiv:0907.1262.

\bibitem{Nunokawa:1996tg}
H.~Nunokawa, Y.~Z. Qian, A.~Rossi and J.~W.~F. Valle,
\newblock Phys. Rev. {\bf D54}, 4356 (1996), [hep-ph/9605301].

\bibitem{Nunokawa:2007qh}
H.~Nunokawa, S.~J. Parke and J.~W.~F. Valle,
\newblock Prog. Part. Nucl. Phys. {\bf 60}, 338 (2008), [arXiv:0710.0554
  [hep-ph]].

\bibitem{Bandyopadhyay:2007kx}
ISS Physics Working Group, A.~Bandyopadhyay {\em et~al.},
\newblock arXiv:0710.4947 [hep-ph].

\bibitem{huber:2001zw}
P.~Huber and J.~W.~F. Valle,
\newblock Phys. Lett. {\bf B523}, 151 (2001), [hep-ph/0108193].

\bibitem{huber:2001de}
P.~Huber, T.~Schwetz and J.~W.~F. Valle,
\newblock Phys. Rev. Lett. {\bf 88}, 101804 (2002), [hep-ph/0111224].

\bibitem{Nunokawa:1996ve}
H.~Nunokawa, A.~Rossi and J.~W.~F. Valle,
\newblock Nucl. Phys. {\bf B482}, 481 (1996), [hep-ph/9606445].

\bibitem{EstebanPretel:2007yu}
A.~Esteban-Pretel, R.~Tom\`as and J.~W.~F. Valle,
\newblock Phys. Rev. {\bf D76}, 053001 (2007), [arXiv:0704.0032 [hep-ph]].

\bibitem{Pastor:2002we}
S.~Pastor and G.~Raffelt,
\newblock Phys. Rev. Lett. {\bf 89}, 191101 (2002), [astro-ph/0207281].

\bibitem{Sawyer:2005jk}
R.~F. Sawyer,
\newblock Phys. Rev. {\bf D72}, 045003 (2005), [hep-ph/0503013].

\bibitem{Fuller:2005ae}
G.~M. Fuller and Y.-Z. Qian,
\newblock Phys. Rev. {\bf D73}, 023004 (2006), [astro-ph/0505240].

\bibitem{Duan:2005cp}
H.~Duan, G.~M. Fuller and Y.-Z. Qian,
\newblock Phys. Rev. {\bf D74}, 123004 (2006), [astro-ph/0511275].

\bibitem{Duan:2006an}
H.~Duan, G.~M. Fuller, J.~Carlson and Y.-Z. Qian,
\newblock Phys. Rev. {\bf D74}, 105014 (2006), [astro-ph/0606616].

\bibitem{Hannestad:2006nj}
S.~Hannestad, G.~G. Raffelt, G.~Sigl and Y.~Y.~Y. Wong,
\newblock Phys. Rev. {\bf D74}, 105010 (2006), [astro-ph/0608695].

\bibitem{Duan:2007mv}
H.~Duan, G.~M. Fuller, J.~Carlson and Y.-Z. Qian,
\newblock Phys. Rev. {\bf D75}, 125005 (2007), [astro-ph/0703776].

\bibitem{Raffelt:2007yz}
G.~G. Raffelt and G.~Sigl,
\newblock Phys. Rev. {\bf D75}, 083002 (2007), [hep-ph/0701182].

\bibitem{EstebanPretel:2007ec}
A.~Esteban-Pretel, S.~Pastor, R.~Tom\`as, G.~G. Raffelt and G.~Sigl,
\newblock Phys. Rev. {\bf D76}, 125018 (2007), [arXiv:0706.2498].

\bibitem{Raffelt:2007cb}
G.~G. Raffelt and A.~Y. Smirnov,
\newblock Phys. Rev. {\bf D76}, 081301 (2007), [arXiv:0705.1830].

\bibitem{Raffelt:2007xt}
G.~G. Raffelt and A.~Y. Smirnov,
\newblock Phys. Rev. {\bf D76}, 125008 (2007), [arXiv:0709.4641].

\bibitem{Duan:2007fw}
H.~Duan, G.~M. Fuller and Y.-Z. Qian,
\newblock Phys. Rev. {\bf D76}, 085013 (2007), [arXiv:0706.4293].

\bibitem{Fogli:2007bk}
G.~L. Fogli, E.~Lisi, A.~Marrone and A.~Mirizzi,
\newblock JCAP {\bf 0712}, 010 (2007), [arXiv:0707.1998].

\bibitem{Duan:2007bt}
H.~Duan, G.~M. Fuller, J.~Carlson and Y.-Q. Zhong,
\newblock Phys. Rev. Lett. {\bf 99}, 241802 (2007), [0707.0290].

\bibitem{Duan:2007sh}
H.~Duan, G.~M. Fuller, J.~Carlson and Y.-Z. Qian,
\newblock Phys. Rev. Lett. {\bf 100}, 021101 (2008), [arXiv:0710.1271].

\bibitem{Dasgupta:2008cd}
B.~Dasgupta, A.~Dighe, A.~Mirizzi and G.~G. Raffelt,
\newblock Phys. Rev. {\bf D77}, 113007 (2008), [0801.1660].

\bibitem{EstebanPretel:2007yq}
A.~Esteban-Pretel, S.~Pastor, R.~Tom\`as, G.~G. Raffelt and G.~Sigl,
\newblock Phys. Rev. {\bf D77}, 065024 (2008), [arXiv:0712.1137].

\bibitem{Dasgupta:2007ws}
B.~Dasgupta and A.~Dighe,
\newblock Phys. Rev. {\bf D77}, 113002 (2008), [arXiv:0712.3798].

\bibitem{Duan:2008za}
H.~Duan, G.~M. Fuller and Y.-Z. Qian,
\newblock Phys. Rev. {\bf D77}, 085016 (2008), [arXiv:0801.1363].

\bibitem{Dasgupta:2008my}
B.~Dasgupta, A.~Dighe and A.~Mirizzi,
\newblock Phys. Rev. Lett. {\bf 101}, 171801 (2008), [arXiv:0802.1481].

\bibitem{Sawyer:2008zs}
R.~F. Sawyer,
\newblock  arXiv:0803.4319.

\bibitem{Duan:2008eb}
H.~Duan, G.~M. Fuller and J.~Carlson,
\newblock Comput. Sci. Dis. {\bf 1}, 015007 (2008), [arXiv:0803.3650].

\bibitem{Chakraborty:2008zp}
S.~Chakraborty, S.~Choubey, B.~Dasgupta and K.~Kar,
\newblock JCAP {\bf 0809}, 013 (2008), [arXiv:0805.3131].

\bibitem{Dasgupta:2008cu}
B.~Dasgupta, A.~Dighe, A.~Mirizzi and G.~G. Raffelt,
\newblock Phys. Rev. {\bf D78}, 033014 (2008), [arXiv:0805.3300].

\bibitem{Fogli:2008fj}
G.~Fogli, E.~Lisi, A.~Marrone and I.~Tamborra,
\newblock JCAP {\bf 0904}, 030 (2009), [arXiv:0812.3031].

\bibitem{Amsler:2008zz}
Particle Data Group, C.~Amsler {\em et~al.},
\newblock Phys. Lett. {\bf B667}, 1 (2008).

\bibitem{Mikheev:1986gs}
  S.~P.~Mikheev and A.~Y.~Smirnov,
  Sov.\ J.\ Nucl.\ Phys.\  {\bf 42}, 913 (1985)
  [Yad.\ Fiz.\  {\bf 42}, 1441 (1985)].

\bibitem{Mikheev:1986wj}
  S.~P.~Mikheev and A.~Y.~Smirnov,
  Nuovo Cim.\  C {\bf 9}, 17 (1986).


\bibitem{Botella:1986wy}
F.~J. Botella, C.~S. Lim and W.~J. Marciano,
\newblock Phys. Rev. {\bf D35}, 896 (1987).

\bibitem{Akhmedov:2002zj}
E.~K. Akhmedov, C.~Lunardini and A.~Y. Smirnov,
\newblock Nucl. Phys. {\bf B643}, 339 (2002), [hep-ph/0204091].

\bibitem{Mirizzi:2009td}
  A.~Mirizzi, S.~Pozzorini, G.~G.~Raffelt and P.~D.~Serpico,
  arXiv:0907.3674 [hep-ph].


\bibitem{Blennow:2008er}
M.~Blennow, A.~Mirizzi and P.~D. Serpico,
\newblock Phys. Rev. {\bf D78}, 113004 (2008), [arXiv:0810.2297].

\bibitem{Nunokawa:1997dp}
H.~Nunokawa, V.~B. Semikoz, A.~Y. Smirnov and J.~W.~F. Valle,
\newblock Nucl. Phys. {\bf B501}, 17 (1997), [hep-ph/9701420].


\bibitem{Gava:2009gt}
  J.~Gava and C.~C.~Jean-Louis,
  arXiv:0907.3947 [hep-ph].

\bibitem{Amanik:2004vm}
P.~S. Amanik, G.~M. Fuller and B.~Grinstein,
\newblock Astropart. Phys. {\bf 24}, 160 (2005), [hep-ph/0407130].

\bibitem{Amanik:2006ad}
  P.~S.~Amanik and G.~M.~Fuller,
  Phys.\ Rev.\  D {\bf 75}, 083008 (2007)
  [arXiv:astro-ph/0606607].

\bibitem{Fornengo:2001pm}
  N.~Fornengo, M.~Maltoni, R.~Tom\`as and J.~W.~F.~Valle,
  Phys.\ Rev.\  D {\bf 65}, 013010 (2002)
  [arXiv:hep-ph/0108043].

\bibitem{Friedland:2004ah}
  A.~Friedland, C.~Lunardini and M.~Maltoni,
  Phys.\ Rev.\  D {\bf 70}, 111301 (2004)
  [arXiv:hep-ph/0408264].

\bibitem{Friedland:2005vy}
  A.~Friedland and C.~Lunardini,
  Phys.\ Rev.\  D {\bf 72}, 053009 (2005)
  [arXiv:hep-ph/0506143].

\bibitem{Escrihuela:2009up}
  F.~J.~Escrihuela, O.~G.~Miranda, M.~A.~Tortola and J.~W.~F.~Valle,
  Phys.\ Rev.\  D {\bf 80}, 105009 (2009)
  [arXiv:0907.2630 [hep-ph]].


\bibitem{Sawyer:2004ai}
R.~F. Sawyer,
\newblock hep-ph/0408265.

\bibitem{Kuo:1989qe}
  T.~K.~Kuo and J.~T.~Pantaleone,
  Rev.\ Mod.\ Phys.\  {\bf 61}, 937 (1989).

\bibitem{Dighe:1999bi}
A.~S. Dighe and A.~Y. Smirnov,
\newblock Phys. Rev. {\bf D62}, 033007 (2000), [hep-ph/9907423].

\bibitem{Fogli:2002xj}
G.~L. Fogli, E.~Lisi, A.~Mirizzi and D.~Montanino,
\newblock Phys. Rev. {\bf D66}, 013009 (2002), [hep-ph/0202269].

\bibitem{EstebanPretel:2008ni}
A.~Esteban-Pretel {\em et~al.},
\newblock Phys. Rev. {\bf D78}, 085012 (2008), [arXiv:0807.0659].

\bibitem{Schirato:2002tg}
R.~C. Schirato, G.~M. Fuller,
\newblock astro-ph/0205390.

\bibitem{Fogli:2003dw}
G.~L. Fogli, E.~Lisi, D.~Montanino and A.~Mirizzi,
\newblock Phys. Rev. {\bf D68}, 033005 (2003), [hep-ph/0304056].

\bibitem{Tomas:2004gr}
R.~Tomas {\em et~al.},
\newblock JCAP {\bf 0409}, 015 (2004), [astro-ph/0407132].

\bibitem{Keil:2002in}
M.~T. Keil, G.~G. Raffelt and H.-T. Janka,
\newblock astro-ph/0208035.

\bibitem{Dasgupta:2009mg}
B.~Dasgupta, A.~Dighe, G.~G. Raffelt and A.~Y. Smirnov,
\newblock Phys. Rev. Lett. {\bf 103}, 051105 (2009), [0904.3542].

\bibitem{Fogli:2006xy}
G.~L. Fogli, E.~Lisi, A.~Mirizzi and D.~Montanino,
\newblock JCAP {\bf 0606}, 012 (2006), [hep-ph/0603033].

\bibitem{Friedland:2006ta}
A.~Friedland and A.~Gruzinov,
\newblock astro-ph/0607244.

\bibitem{Gava:2009pj}
J.~Gava, J.~Kneller, C.~Volpe and G.~C. McLaughlin,
\newblock Phys. Rev. Lett. {\bf 103}, 071101 (2009), [arXiv:0902.0317].

\bibitem{Kachelriess:2004ds}
M.~Kachelriess {\em et~al.},
\newblock Phys. Rev. {\bf D71}, 063003 (2005), [astro-ph/0412082].

\bibitem{Lunardini:2001pb}
C.~Lunardini and A.~Y. Smirnov,
\newblock Nucl. Phys. {\bf B616}, 307 (2001), [hep-ph/0106149].

\bibitem{Dighe:2003be}
A.~S. Dighe, M.~T. Keil and G.~G. Raffelt,
\newblock JCAP {\bf 0306}, 005 (2003), [hep-ph/0303210].

\bibitem{Dighe:2003jg}
A.~S. Dighe, M.~T. Keil and G.~G. Raffelt,
\newblock JCAP {\bf 0306}, 006 (2003), [hep-ph/0304150].

\bibitem{Dighe:2003vm}
A.~S. Dighe, M.~Kachelriess, G.~G. Raffelt and R.~Tomas,
\newblock JCAP {\bf 0401}, 004 (2004), [hep-ph/0311172].

\end{thebibliography}

\end{document}